\documentclass[runningheads]{llncs}
\newif\ifshort
\shortfalse

\long\def\@makecaption#1#2{}
\usepackage{graphicx}
\usepackage{subcaption}
\usepackage{xspace}
\usepackage{listings,array,varwidth}
\usepackage{courier}
\usepackage{makecell}
\usepackage{svg}
\usepackage{booktabs}
\usepackage{hyperref}
\usepackage[capitalise]{cleveref}
\crefname{algorithm}{Alg.}{Algs.}
\crefname{section}{Sec.}{Secs.}
\crefname{definition}{Def.}{Defs.}
\crefname{table}{Tab.}{Tabs.}
\crefname{appendix}{App.}{Apps.}

\newcommand{\awsccommon}{\texttt{aws-c-common}\xspace}
\newcommand{\cbmc}{\textsc{CBMC}\xspace}
\newcommand{\seahorn}{\textsc{SeaHorn}\xspace}
\newcommand{\klee}{\textsc{KLEE}\xspace}
\newcommand{\libfuzzer}{\texttt{libFuzzer}\xspace}

\newcommand{\CaS}{CaS\xspace}

\usepackage[colorinlistoftodos]{todonotes} \newcommand{\yv}[1]{\textcolor{orange}{[*YV: #1*]}\xspace}

\newcommand{\ys}[1]{\textcolor{blue}{[YS: #1]}\xspace}
\newcommand{\xz}[1]{\textcolor{red}{[XZ: #1]}\xspace}

\renewcommand{\yv}[1]{}
\renewcommand{\ys}[1]{}
\renewcommand{\xz}[1]{}

\newcommand{\code}[1]{\lstinline[basicstyle=\footnotesize\ttfamily] {#1}\xspace}

\newcommand{\bparagraph}[1]{\noindent\textbf{#1}}

\begin{document}
\title{Verifying Verified Code\thanks{This research was supported by grants
    from WHJIL and NSERC CRDPJ 543583-19.}}
\author{Siddharth Priya\inst{1}
  \and Xiang Zhou\inst{1}
  \and Yusen Su\inst{1}
  \and Yakir Vizel\inst{2}
  \and Yuyan Bao\inst{1}
  \and Arie Gurfinkel\inst{1}
}
\authorrunning{S. Priya et al.}
\institute{University of Waterloo\inst{1} and The Technion\inst{2}}
\maketitle              \begin{abstract}
A recent case study from AWS by Chong et al. proposes an effective methodology for Bounded Model Checking in industry.  In this paper, we report on a followup case study that explores the methodology from the perspective of three research questions: (a) can proof artifacts be used across verification tools; (b) are there bugs in verified code; and (c) can specifications be improved. To study these questions, we port the verification tasks for \awsccommon library to \seahorn and \klee. We show the benefits of using compiler semantics and cross-checking specifications with different verification techniques, and call for standardizing proof library extensions to increase specification reuse. The verification tasks discussed are publicly available online.
\end{abstract} \section{Introduction}
\label{sec:agintro}
Bounded Model Checking (BMC) is an effective static analysis technique that
reduces program analysis to propositional satisfiability (SAT) or Satisfiability Modulo
Theories (SMT). It works directly on the source code. It is very precise, e.g.,
accounting for semantics of the programming language, memory models, and
machine arithmetic.
There is a vibrant ecosystem of tools from academia (e.g.,
SMACK~\cite{DBLP:conf/cav/RakamaricE14},
CPAChecker~\cite{DBLP:conf/cav/BeyerK11},
ESBMC~\cite{DBLP:conf/kbse/GadelhaMMC0N18}), industrial research labs (e.g.,
Corral~\cite{DBLP:conf/sigsoft/LalQ14},
F-SOFT~\cite{DBLP:conf/cav/IvancicYGGSA05}), and industry (e.g.,
CBMC~\cite{DBLP:conf/tacas/ClarkeKL04}, Crux~\cite{Manual:crux}, QPR~\cite{DBLP:conf/vstte/BuningSF20}). There is an
annual software verification competition, SV-COMP~\cite{DBLP:conf/tacas/Beyer20}, with many
participants. However, with a few exceptions, BMC is not actively used in
software industry. Especially, when compared to dynamic analysis techniques such as fuzzing~\cite{Manual:libfuzzer}, or light-weight formal methods such as static analysis \cite{DBLP:journals/cacm/BesseyBCCFHHKME10}.

Transitioning research tools into practice requires case-studies, methodology,
and best-practices to show how the tools are best applied. Until recently,
there was no publicly available industrial case study on successful application
of BMC for continuous verification\footnote{By \emph{continuous verification}, we mean verification that is integrated with continuous integration (CI) and is checked during every commit.} of C code. This has changed
with~\cite{DBLP:conf/icse/ChongCKKMSTTT20} -- a case study from the Automated
Reasoning Group (ARG) at Amazon Web Services (AWS) on the use of CBMC for
proving memory safety (and other properties) of several AWS C libraries. This
case study proposes a verification methodology with two core principles:
(a) verification tasks structured around units of functionality (i.e., around a
single function, as in a unit test), and (b) the use of code 
to express specifications (i.e., pre-, post-conditions,
and other contextual assumptions). We refer to these as
\emph{unit proofs}, and \emph{Code as Specification (CaS)}, respectively. The
methodology is efficient because small verification tasks help alleviate
scalability issues inherent in BMC. 
More significantly, developers adopt, own, extend and even
use specifications (as code) in other contexts, e.g., unit tests. 
Admirably, AWS has released all of the verification artifacts (code, specifications and
verification libraries)\footnote{\url{https://github.com/awslabs/aws-c-common/tree/main/verification/cbmc}}.
Moreover, these are maintained and integrated
into Continuous Integration (CI). 
This gave us a unique opportunity to study, validate, and refine the methodology
of~\cite{DBLP:conf/icse/ChongCKKMSTTT20}.
In this paper, we report on our experience on 
adapting the verification tasks of~\cite{DBLP:conf/icse/ChongCKKMSTTT20} to two new verification tools: a Bounded Model Checking engine of \seahorn, and the symbolic execution tool \klee.
We present our experience as a case study that is organized around three Research Questions (RQ):

\paragraph{RQ1: Does CaS empower multiple tools for a common verification task?} 
Code is the \emph{lingua franca} among developers, compilers, and verification
tools. Thus, \CaS makes specifications understandable by
multiple verification tools. 
To validate effectiveness of this hypothesis, we adapted the unit proofs from
AWS to different tools, and report on the experience in Sec.~\ref{sec:rq1}.
While giving a positive answer to RQ1, we highlight the importance of the
semantics used to interpret \CaS, and that effectiveness of each tool depends on
specification styles.

\paragraph{RQ2: Are there bugs in verified  code?}
Specifications written by humans may have errors.
Do such errors hide bugs in verified implementations? 
What sanity checks are helpful to find bugs in implementations \emph{and}
specifications? The public availability of~\cite{DBLP:conf/icse/ChongCKKMSTTT20}
is a unique opportunity to study this question. In contrast
to~\cite{DBLP:conf/icse/ChongCKKMSTTT20}, we found no new bugs in
the library being verified (\texttt{aws-c-common}). However, we have found multiple errors in specifications!
Reporting them to AWS triggered a massive review of existing unit proofs with
many similar issues found and fixed. We report the bugs, and 
techniques that helped us discover them, in Sec.~\ref{sec:rq2}.

\paragraph{RQ3: Can specifications be improved while maintaining \CaS philosophy?}
Some mistakes in specifications can be prevented by improvements to the specification
language. 
We propose a series of improvements that significantly reduce specification burden.
They are mostly in the form of built-in functions, thus, familiar to developers.
In particular, we show how to make the verification of the \texttt{linked\_list} data
structure in \texttt{aws-c-common} significantly more efficient,
while making the proofs unbounded (i.e., correct
for linked list of any size).

In our case study, we used the BMC engine of
\seahorn~\cite{DBLP:conf/cav/GurfinkelKKN15} and symbolic execution tool
\klee~\cite{DBLP:conf/osdi/CadarDE08}. We have chosen \seahorn because it is conceptually similar to CBMC that was used in~\cite{DBLP:conf/icse/ChongCKKMSTTT20}. Thus, it was reasonable to assume that all verification tasks can be ported to it. We are also intimately familiar with \seahorn. Thus, we did not only port verification tasks, but proposed improvements to \seahorn to facilitate the process. We have chosen \klee because it is a well-known representative of symbolic execution -- an approach that is the closest alternative to Bounded Model Checking. 

Overall, we have ported all of the $169$~unit proofs  of \texttt{aws-c-common} to \seahorn, and $153$~to \klee. The case study represents a year of effort.
The time was divided between porting verification tasks, improving \seahorn to allow for a better comparison, and, many manual and semi-automated sanity checks to increase confidence in specifications.
Additionally, we have experimented with using unit proofs as fuzz targets using LLVM fuzzing library \libfuzzer~\cite{Manual:libfuzzer} and adapted $146$ of the unit proofs to \libfuzzer.

We make all results of our work publicly available and reproducible at \url{https://github.com/seahorn/verify-c-common}. In addition to what is reported in this paper, we have developed an extensive \textsc{CMake} build system that simplifies integration of additional tools. The case study is \emph{live} in the sense that it is integrated in CI and is automatically re-run nightly. Thus, it is synchronized both with the tools we use and the AWS library we verify.

We hope that our study inspires researchers
to adapt their tools to industrial code, and inspires industry to 
release verification efforts to study.

\paragraph{Caveats and non-goals.} We focus on the issues of methodology and sharing verification tasks between different tools. The tools that we use have different strengths and weaknesses.  While they all validate user-supplied assertions, they check for different built-in properties (e.g., numeric overflow, undefined behaviours, memory safety). The goal is not to compare the tools head-to-head, or to find the best tool for a given task. We have not attempted to account for the differences between the tools. Nor have we tried to completely cover all verification tasks by all tools. Our goal was to preserve the unit proofs of~\cite{DBLP:conf/icse/ChongCKKMSTTT20} as much as possible to allow for a better comparison. For that reason, while we do report on performance results for the different tools, we do not describe them in detail. An interested reader is encouraged to look at the detailed data we make available on GitHub. Furthermore, while we have applied fuzzing to the unit proofs, we do not focus on effectiveness and applicability of static vs dynamic verification but only on the issues of methodology.

To summarize, we make the following contributions: (a) we validate that \CaS can be used to share specifications between multiple tools, especially tools that share the same techniques (i.e., BMC), or tools with related techniques (i.e., BMC and Symbolic Execution); (b) we describe in details bugs that are found in verified code (more specifically, in specifications), some are quite surprising; (c) we suggest a direction to improve \CaS with additional built-in functions that simplify common specification; and (d) we make our system publicly available allowing other researches to integrate their tools,  use it as a benchmark, and to validate new verification approaches on industrial code.

The rest of the paper is structured as follows. \cref{sec:background} recalls the methodology of unit proof and  \CaS. And \cref{sec:questions} presents the architecture of the case study and answers the three research questions. We discuss related work in \cref{sec:related} and offer concluding remarks in \cref{sec:conclusion}.

 \section{Unit Proofs with Code-as-Specification}
\label{sec:background}

In~\cite{DBLP:conf/icse/ChongCKKMSTTT20}, a methodology for program verification
is proposed that allows developers to write specifications and proofs using the
C programming language.
The core of the methodology are \emph{unit proof}\footnote{In~\cite{DBLP:conf/icse/ChongCKKMSTTT20}, these are called
  \emph{proof harnesses}.} and \emph{Code as Specification (\CaS)}.
  A \emph{unit proof} is similar to a \emph{unit test} in that it is
a piece of a code (usually a method) that invokes another piece of code (under
test) and checks its correctness~\cite{osherove2009art}. 
  \cref{fig:harness-get-ptr} shows an example of a
unit proof for the method \code{aws\_array\_list\_get\_at\_ptr}, from
\awsccommon library. It has three parts: (1) the specification of
\code{aws\_array\_list\_get\_at\_ptr}, i.e., pre- (line 8) and
post-conditions (lines 10--11); (2) a call to the function under verification
(line~9); and (3) the specification of the program context that the method is
called from (lines 2--7). Note that all specifications are written directly in C.  We call this specification style -- \CaS. 
Assumptions (or pre-conditions) correspond to 
\code{\_\_CPROVER\_assume}, and assertions (or post-conditions) correspond to 
\code{assert}. Specifications are factored into functions. For
example, \code{aws\_array\_list\_is\_valid} specifies a
representation invariant of the array list. In this unit proof, the context is
restricted to a list of bounded size but with unconstrained elements and an \code{index}
with (intentionally) unspecified value of type \code{size\_t}. Even without
expanding the code further, its meaning is clear to any C
developer familiar with the library.

The unit proof is verified with \cbmc~\cite{DBLP:conf/tacas/ClarkeKL04}. \cbmc
uses a custom SMT solver to check that there are no executions that satisfy the
pre-conditions and violate at least one of the assertions (i.e., a
counterexample). Together with the explicit assertions, \cbmc checks built-in properties: \mbox{memory safety and integer overflow.}

\newsavebox{\figonebox}
\begin{lrbox}{\figonebox}\begin{lstlisting}
void aws_array_list_get_at_ptr_harness() {
  struct aws_array_list list;
  /* memhavoc(&list, sizeof(struct aws_array_list))); */
  __CPROVER_assume(aws_array_list_is_bounded(&list));
  ensure_array_list_has_allocated_data_member(&list);
  void **val = can_fail_malloc(sizeof(void *));
  size_t index /* = nd_size_t() */;
  __CPROVER_assume(aws_array_list_is_valid(&list) && val != NULL);
  if (aws_array_list_get_at_ptr(&list, val, index) == AWS_OP_SUCCESS)
    assert(list.data != NULL && index < list.length);
  assert(aws_array_list_is_valid(&list)); }
\end{lstlisting}
\end{lrbox}\begin{figure}[t]
\centering
\scalebox{0.8}{\usebox{\figonebox}}
\caption{The unit proof of \texttt{aws\_array\_list\_get\_at\_ptr} from \cite{DBLP:conf/icse/ChongCKKMSTTT20}.} 
\label{fig:harness-get-ptr}
\end{figure}

According to~\cite{DBLP:conf/icse/ChongCKKMSTTT20},  CaS and unit proofs are a practical and productive verification methodology. It has been used
successfully to verify memory safety (and other properties) of multiple AWS projects, including the \awsccommon library that we use in our case study. The library provides cross platform configuration, data
structures, and error handling support to a range of other AWS C libraries and
SDKs. It is the foundation of many security related libraries, such as the AWS
Encryption SDK for C~\cite{DBLP:conf/icse/ChongCKKMSTTT20}.
It contains $13$ data structures, $169$ unit proofs that verify over $20$K lines of code (LOC). \cref{tab:time-comp} shows the LOC and running time for each data structure.

\begin{figure}[tb]
    \centering
    \scalebox{0.8}{\begin{tabular}{p{2cm} p{0.8cm} p{0.8cm} p{0.8cm} p{0.8cm} p{0.1cm} p{1cm} p{0.8cm} p{0.1cm} | p{0.8cm} p{0.6cm} p{0.1cm} p{0.8cm} p{0.8cm}  c}
         \toprule
         & & \multicolumn{3}{c}{LOC} & & \multicolumn{2}{c}{\cbmc (s)} & & \multicolumn{2}{c}{\seahorn (s)} & & \multicolumn{3}{c}{\klee (s)} \\
         \cmidrule{3-5} \cmidrule{7-8} \cmidrule{10-11}
        category & num & avg & min & max & & avg & std & & avg & std & & count & avg & std \\
        \midrule
        arithmetic & 6 & 33 & 11 & 40 &  & 3.8 & 0.8 &  & 0.6 & 0.1 &  & 6 & 0.9 & 0.3  \\
        array & 4 & 97 & 78 & 112 &  & 5.6 & 0.0 &  & 1.7 & 0.7 &  & 4 & 32.3 & 6.0  \\
        array\_list & 23 & 126 & 77 & 181 &  & 35.8 & 60.8 &  & 2.5 & 3.3 &  & 23 & 55.4 & 49.3  \\
        byte\_buf & 29 & 97 & 50 & 188 &  & 17.6 & 47.3 &  & 1.0 & 0.8 &  & 27 & 75.3 & 124.1  \\
        byte\_cursor & 24 & 98 & 47 & 179 &  & 6.9 & 3.8 &  & 1.0 & 0.5 &  & 17 & 12.8 & 14.4  \\
        hash\_callback & 3 & 115 & 49 & 198 &  & 9.7 & 5.5 &  & 4.9 & 3.6 &  & 3 & 64.0 & 45.5  \\
        hash\_iter & 4 & 177 & 169 & 185 &  & 12.8 & 9.2 &  & 9.2 & 15.0 &  & 3 & 20.8 & 9.7  \\
        hash\_table & 19 & 172 & 36 & 328 &  & 23.5 & 33.3 &  & 5.3 & 7.5 &  & 15 & 104.6 & 333.4  \\
        linked\_list & 18 & 115 & 17 & 219 &  & 58.9 & 209.4 &  & 2.0 & 2.1 &  & 18 & 0.7 & 0.1  \\
        others & 2 & 15 & 10 & 21 &  & 3.5 & 0.0 &  & 0.5 & 0.0 &  & 1 & 0.7 & --  \\
        priority\_queue & 15 & 187 & 136 & 258 &  & 208.1 & 303.4 &  & 10.6 & 16.9 &  & 15 & 46.4 & 11.6  \\
        ring\_buffer & 6 & 155 & 56 & 227 &  & 20.0 & 19.5 &  & 29.5 & 34.2 &  & 6 & 48.1 & 26.4  \\
        string & 15 & 87 & 11 & 209 &  & 6.3 & 1.3 &  & 2.9 & 1.8 &  & 15 & 139.7 & 159.7  \\
        \midrule
        \textbf{Total} & 168 & \textsc{Loc} & 20,190 & & & \textsc{Time} & 6,475 & & \textsc{Time} & 691 & & \textsc{Time} & 8,577 \\
        \bottomrule
    \end{tabular}}
    \caption{Verification results for \cbmc, \seahorn and \klee.}
    \label{tab:time-comp}
\end{figure}

 \section{Case Study}
\label{sec:questions}
\vspace{-0.1in}
\begin{figure}[t]
  \centering
  \includegraphics[scale=0.45]{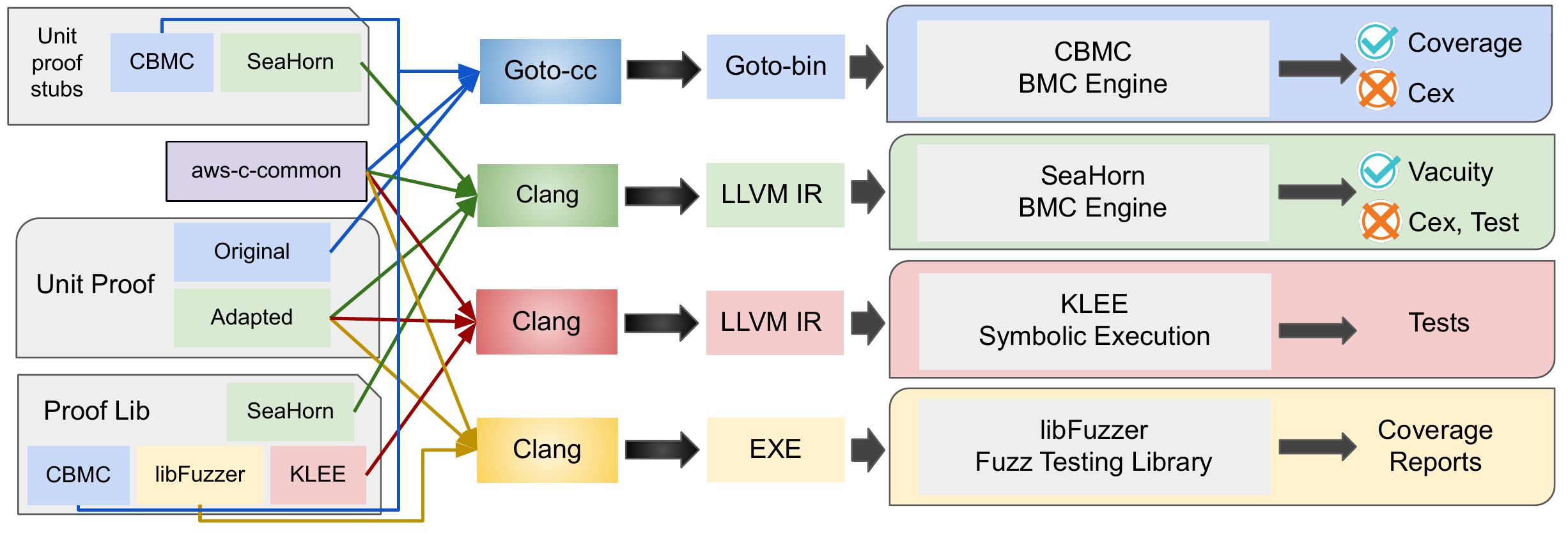}
  \caption{Architecture of the case study.}
  \label{fig:arch}
\end{figure}

The architecture of our case study is shown
in~\cref{fig:arch}. To compare with \cbmc, we use two tools based on the LLVM
framework~\cite{DBLP:conf/cgo/LattnerA04}: \seahorn and \klee.
\seahorn~\cite{DBLP:conf/cav/GurfinkelKKN15} is a verification
framework. We used the bit- and memory-precise BMC 
developed during the case study. Its techniques are
closest to CBMC.
\klee~\cite{DBLP:conf/osdi/CadarDE08} is a well-known symbolic execution tool.
It is an alternative to BMC for bounded exhaustive verification.
In addition, we have experimented with \libfuzzer\ -- a coverage-guided random testing framework. It does no symbolic
reasoning, and, together with address sanitizer, is known to be effective at
discovering memory errors. Fuzzing results are available online. \footnote{\url{https://seahorn.github.io/verify-c-common/fuzzing_coverage/index.html}.}

The rest of the section describes the research questions and our findings.

 \subsection{RQ1: Does CaS Empower Multiple Tools?}
\label{sec:rq1}

Hypothetically, \CaS methodology enables sharing the same formal specification among multiple,
potentially distinct, tools and techniques. For example, semantic analyses of
IDEs and compilers can catch simple semantics bugs and inconsistencies
in specifications. Fuzzers can validate specifications through  testing.
Symbolic execution can supplement BMC by capitalizing on a different balance in
performance versus precision. Static analysis tools can be used to compute
inductive invariants. However, is the hypothesis true in practice? 

To validate the hypothesis, we adapted the unit proofs from \awsccommon to
two distinct verification techniques: BMC with \seahorn and symbolic execution
with \klee. We have also attempted to use unit proofs as fuzz targets for \libfuzzer. 
While our experience
supports the hypothesis, we encountered two major challenges:
\emph{semantics} and \emph{effectiveness of specifications}.

\paragraph{Semantics.} Code without semantics
is meaningless. Developers understand code without being versed in formal
semantics, however, many technical details and ``corner cases''
are often debated. This is especially true for C -- \emph{``the
  semantics of C has been a vexed question for much of the last 40
  years''}~\cite{DBLP:conf/pldi/MemarianMLNCWS16}. Clear semantics are crucial
when code (and \CaS) are used with multiple tools. 

The unit proofs in~\cite{DBLP:conf/icse/ChongCKKMSTTT20} do not follow the C
semantics. For example, consider the proof in \cref{fig:harness-get-ptr}.
According to C, it has no meaning as both \code{list} (line~2) and \code{index}
(line~7) are used uninitialized. CBMC treats uninitialized variables as
non-deterministic. So it is well-defined for CBMC, but not for other tools.

What is a good choice of semantics for \CaS?
In~\cite{DBLP:conf/pldi/MemarianMLNCWS16}, two semantics are described -- the
ISO C Standard and the \emph{de facto} semantics of compilers.
Developers understand (and use) the de facto semantics. For example, comparison of
arbitrary pointers is undefined according to ISO C, but defined consistently in
mainstream compilers (and used in \awsccommon!). Therefore, we argue
that \CaS must use the de facto semantics. Furthermore, unit proofs must
be compilable and, therefore, executable, so developers can execute them not \emph{just} in their heads (like with~\cite{DBLP:conf/icse/ChongCKKMSTTT20}). Note that de facto semantics is not complete with regards to C 
semantics, but is a commonly agreed upon subset. What de facto semantics does not cover is compiler dependent semantics.
 
In our experience, using \CaS with the de facto semantics is not hard. For
example, to adapt \cref{fig:harness-get-ptr}, we introduced \code{memhavoc} and
\code{nd_size_t}, shown as comments, that fills a memory region at a given
address with non-deterministic bytes, and returns a non-deterministic value of
type \code{size_t}, respectively.

\paragraph{Effectiveness of specifications.} 
We used three different tools on the same unit proof.
Each tool requires slightly different styles of specifications to be effective. We believe that these
stylistic differences between specifications can be captured by traditional
code refactoring techniques (i.e., functions, macros, etc.).
However, this is not easy whenever the specifications have not been written with
multiple tools (and with their strengths and weaknesses) in mind. A significant part of our work has been in refactoring unit proofs from~\cite{DBLP:conf/icse/ChongCKKMSTTT20} to be more modular.

We illustrate this with the pre-condition for the \code{byte_buf} data-structure. In~\cite{DBLP:conf/icse/ChongCKKMSTTT20}, data structures are assumed to be initially non-deterministic, and various assumptions throughout the unit proof are used to restrict it (e.g., lines~2--5 in \cref{fig:harness-get-ptr}). 
This impedes specification re-use since different tools work well with different styles of pre-conditions. For example, symbolic execution and fuzzing require memory to be explicitly allocated, and all tools that use de-facto semantics require all memory be initialized before use.

For \code{byte_buf}, we factored out its pre-conditions into a function \code{init_byte_buf}.\footnote{Similarly, we introduced \code{init_array_list} to replace lines~2--5 in \cref{fig:harness-get-ptr}.} Its implementations for \seahorn, \klee, and \libfuzzer are shown in \cref{fig:init-impls}. It takes \code{buf} structure as input, and initializes its fields to be consistent with the representation invariant of \code{byte_buf}.

\newsavebox{\figseainitlstbox}
\begin{lrbox}{\figseainitlstbox}\begin{lstlisting}[escapechar=@]
    size_t len = nd_size_t();
    size_t cap = nd_size_t();
    assume(len <= cap);
    assume(cap <= MAX_BUFFER);

    buf->len = len;
    buf->capacity = cap;
    buf->buffer = can_fail_malloc(
      cap * sizeof(*(buf->buffer)));
    buf->allocator = sea_allocator();
  
  
  
  
    @ @
  \end{lstlisting}
\end{lrbox}

\newsavebox{\figkleeinit}
\begin{lrbox}{\figkleeinit}\begin{lstlisting}[numbers=none]
    size_t cap = nd_size_t();
    assume(cap <= MAX_BUFFER);
    buf->buffer = can_fail_malloc(
      cap * sizeof(*(buf->buffer)));
    if (buf->buffer) {
        size_t len = nd_size_t();
        assume(len <= cap);
        buf->len = len;
        buf->capacity = cap;
    }
    else {
        buf->len = 0;
        buf->capacity = 0;
    }
    buf->allocator = sea_allocator();
  \end{lstlisting}
\end{lrbox}

\newsavebox{\figfuzzinit}
\begin{lrbox}{\figfuzzinit}\begin{lstlisting}[escapechar=@,numbers=none]
      size_t len = nd_size_t();
      size_t cap = nd_size_t();@\label{line:libfuzz-cap}@
      cap len = (cap == 0) ? 0 : len 

      buf->len = len;
      buf->capacity = cap;
      buf->buffer = can_fail_malloc(
        cap * sizeof(*(buf->buffer)));
      buf->allocator = sea_allocator();



      @ @
  \end{lstlisting}
\end{lrbox}

\begin{figure}[t]
  \centering
  \begin{subfigure}[t]{0.3\textwidth}
  \scalebox{0.7}{\usebox{\figseainitlstbox}}
  \caption{for \seahorn} 
  \label{fig:init-sea}
  \end{subfigure}
  \hfill  \begin{subfigure}[t]{0.3\textwidth}
  \scalebox{0.7}{\usebox{\figkleeinit}}
  \caption{for \klee} 
  \label{fig:init-klee}
  \end{subfigure}
  \hfill
  \begin{subfigure}[t]{0.3\textwidth}
\scalebox{0.7}{\usebox{\figfuzzinit}}
  \caption{for \libfuzzer} 
  \label{fig:init-fuzz}
  \end{subfigure}
  \caption{Tool-specific implementations for \code{initialize_byte_buf}.} 
  \label{fig:init-impls}
  \end{figure}

\vspace{0.1in}
\bparagraph{\seahorn} initialization is closest to the original of~\cite{DBLP:conf/icse/ChongCKKMSTTT20}. Fields are initialized via calls to external functions (\code{nd_<type>}) that are assumed to return arbitrary values. Representation invariants (i.e., length is less or equal to capacity), as well as any upper bounds on buffer size are specified with \emph{assumptions}. Note that \code{can_fail_malloc} internally initializes allocated memory via a call to \code{memhavoc}, ensuring that reading \code{buf->buffer} is well-defined.   

\vspace{0.1in}
\bparagraph{\klee} initialization is similar to \seahorn, but special care must be taken about the placement of assumptions, and implementation of \code{can_fail_malloc}. In particular, \klee prefers that memory allocation functions are given explicit size, otherwise, it picks a concrete size non-deterministically. Special cases, like \code{buf->buffer} being \code{NULL}, are split in the initialization to aid \klee during symbolic execution. Similar changes can be done for \seahorn, but are not as effective. For that reason, we chose to keep \seahorn initialization as close to~\cite{DBLP:conf/icse/ChongCKKMSTTT20} as possible, but adjusted the one for \klee to be most effective.

\vspace{0.1in}
\bparagraph{\libfuzzer} initialization is the  most different since non-determinisim must be replaced by randomness. In this case, \code{nd_<type>} 
functions are implemented using the random inputs generated by \libfuzzer. 
\emph{Assumptions} are implemented by aborting the current 
fuzzing run if the condition evaluates to \code{false}. Of course, this limits fuzzing effectiveness since the fuzzer must randomly guess inputs to pass all of the assumptions. For that reason, as many assumptions as possible are modeled by an explicit initialization. For example, in line~3 of \cref{fig:init-fuzz}, 
\code{cap} is re-assigned to the modulo of \code{MAX_BUFFER} if \libfuzzer generated 
a value exceeding \code{MAX_BUFFER}. This way, code after line~3 always executes 
regardless of the return value of \code{nd_size_t()} in  line~\ref{line:libfuzz-cap}.

Overall, our results indicate that \CaS empowers multiple verification tools to share 
specifications among them. Common refactoring techniques make
specifications sharing effective. Specifications are easiest to share among tools 
that use similar techniques. 

\paragraph{Discussion.} We conclude this section with a discussion of our experience in using de facto semantics.
First, the code of \awsccommon is written with de facto semantics in mind. We found
that in~\cite{DBLP:conf/icse/ChongCKKMSTTT20} it had to be extended with many conditional compilation flags to provide alternative implementations that are compatible with CBMC or that instruct CBMC to ignore some seemingly undefined behavior. However, we have not changed any lines of \awsccommon. We analyze the
code exactly how it is given to the compiler -- improving 
coverage.
Second, a compiler may generate different target code for different
architectures. 
By using the compiler as front-end, we check that the code is
correct as compiled on different platforms. This is another advantage of
\CaS.
Third, compilers may provide additional safety checks.
For example, \awsccommon uses GCC/Clang
built-in functions for overflow-aware arithmetic. By using de facto semantics, all the
tools used in the case study were able to deal with this in both \CaS and code seamlessly.
Fourth, \awsccommon uses inline assembly to deal with speculative execution-based vulnerabilities~\cite{DBLP:tr/meltdown/s18}. While inline
assembly is not part of the ISO C standard, it is supported by 
compilers. Thus, it is not a problem for \libfuzzer. We developed techniques to handle inline asm in \seahorn. For \klee, we had to ignore such unit proofs.

 \subsection{Are there bugs in verified  code?}
\label{sec:rq2}

Specifications may have errors as they are just programs: ``Writing
specifications can be as error-prone as writing
programs''.\cite{goog:moy2010tokeneer}
Although~\cite{DBLP:conf/icse/ChongCKKMSTTT20} suggests to use code coverage and
code review to increase the confidence in specifications, we still found
non-trivial bugs. We summarize three most interesting ones below.
\ifshort\else A complete list of all bugs we found is available in~\cref{sec:bugs}.\fi

\newsavebox{\figsevenbox}
\begin{lrbox}{\figsevenbox}\begin{lstlisting}[escapechar=@]
typedef 
struct byte_buf { 
  char* buf; 
  int len, cap; 
} BB;
bool BB_is_ok(BB *b) 
{ return (b->len == 0@\label{line:bug1-lencap}@
        || b->buf);  } 
\end{lstlisting}
\end{lrbox}

\newsavebox{\figeightbox}
\begin{lrbox}{\figeightbox}\begin{lstlisting}[escapechar=@]
  assume(0 <= b && b <= 10);
  if (a < (b - 5) &&@\label{line:bug2-ifcond1}@
      a >= (b + 5))@\label{line:bug2-ifcond2}@
  {
    assert(c > 0);@\label{line:bug2-assert}@
  } 
  
  
@ @
\end{lstlisting}
\end{lrbox}

\newsavebox{\figninebox}
\begin{lrbox}{\figninebox}\begin{lstlisting}[escapechar=@]
void ht_del_over(HASH_TB *t) {
  /* remove entry */
  /* t->entry_count--; */@\label{line:bug3-count}@
}



@ @
\end{lstlisting}
\end{lrbox}

\begin{figure}[t]
\centering
\begin{subfigure}[t]{0.3\textwidth}
\scalebox{0.8}{\usebox{\figsevenbox}}
\caption{bug 1} 
\label{fig:bug1}
\end{subfigure}
\hfill
\begin{subfigure}[t]{0.3\textwidth}
\scalebox{0.8}{\usebox{\figeightbox}}
\caption{bug 2} 
\label{fig:bug2}
\end{subfigure}
\hfill
\begin{subfigure}[t]{0.3\textwidth}
\scalebox{0.8}{\usebox{\figninebox}}
\caption{bug 3} 
\label{fig:bug3}
\end{subfigure}
\caption{Simplified code for specification bugs.} 
\label{fig:bug_123}
\end{figure}

\paragraph{Bug 1.}  \cref{fig:bug1} shows the definition of \code{byte buffer} that is a length delimited byte string. Its data representation should be either the buffer (\code{buf}) is \code{NULL} or its capacity (\code{cap}) is 0 (not the \code{len} as defined in ~\code{BB_is_ok}). 
We found this bug by a combination of sanity checks in \seahorn and our model of
the memory allocator (i.e., \code{malloc}). \ifshort\else More details are explained in~\cref{sec:bug-byte_buf}.\fi The bug did not manifest in~\cite{DBLP:conf/icse/ChongCKKMSTTT20} because other pre-conditions ensured that \code{buf} is always allocated. Our report of this bug to AWS triggered a massive code auditing effort in \awsccommon and related libraries where many related issues were found.\footnote{An example is  \url{https://github.com/awslabs/aws-c-common/pull/686/commits}.}

\paragraph{Bug 2.}\label{para:rq2-bug2} 
\cref{fig:bug2} shows a verification pattern where a property (line 5) is checked on the program path (from lines~1 to 5). As the condition at lines 2 and 3 can never be true, the property cannot be checked either.
Our vacuity detection (discussed later) found the bugs occurring in this
pattern. \ifshort\else More details on the bug can be found
in~\cref{sec:pq_s_swap}. \fi Note that the bug was missed by the code coverage detection used by \cbmc, thus, may have been present for several years.

\paragraph{Bug 3.}\label{para:rq2-bug3} 
To make verification scalable, the verification of method A that calls another method B may use a \emph{specification stub} that approximates the functionality of B.
AWS adopts this methodology when verifying the iterator of a hash table. The
iterator calls a function \code{ht_del} to remove an element in a hash table.
During verification \code{ht_del} is approximated by a specification stub shown
in ~\cref{fig:bug3}. However, the approximation does not decrement
\code{entry_count}, i.e., line~\ref{line:bug3-count} should be added to the spec
for correct behavior. In~\cite{DBLP:conf/icse/ChongCKKMSTTT20}, the use of the
buggy stub hides an error in the specification. \ifshort\else See more details
in ~\cref{sec:bug-hash_iter}. \fi

\paragraph{Discussion.} 
Code coverage of a unit proof is, at best, a sanity check for \CaS. It reports which source lines of the specification and code under verification are covered under execution.  However, because source lines can remain uncovered for legitimate reasons e.g., dead code, interpreting a coverage report is not straightforward. There is no obvious \emph{pass/fail} criterion. Thus, we found that code coverage may be insufficient to detect bugs in CaS reliably. 
In fact, bugs exist for multiple years even after code coverage failures. To help find bugs in \CaS with \seahorn, we adapted \emph{vacuity detection}~\cite{DBLP:conf/concur/Kupferman06} to detect unreachable post-conditions.  Vacuity detection checks that every assert statement is reachable.
We encountered engineering challenges when developing vacuity detection.
For example, we received spurious warnings due to code duplication. We silenced such warnings by only reporting a warning if all duplicate asserts reported a vacuity failure. 
In addition, due to \CaS, an unreachable assertions may be removed by compiler's dead code elimination. This is not desirable for vacuity detection. To mitigate this issue, we report when dead code is eliminated. However, since many eliminations are unrelated to specs, there is noise in the report which makes it un-actionable. Interaction between dead code removal by the compiler and vacuity detection remains an open challenge for us.

We have found bugs in specifications, but we do not know what bugs remain. As shown in this section, the bugs were found with a combination of manual auditing and tools. However, these  techniques are far from complete.

 \subsection{Can specifications be improved while maintaining the \CaS philosophy?}
\label{sec:rq3}

\newsavebox{\figtenbox}
\begin{lrbox}{\figtenbox}\begin{lstlisting}[escapechar=@]
linked_list l;@\label{line:rq3_orig_pre_start}@
Node *p = malloc(sizeof(Node));
l.head.next = p;
for (int idx=0; idx < MAX; idx++) {@\label{line:rq3_orig_create}@   
  Node *n = malloc(sizeof(Node)); 
  p->next = n;
  p = n; }@\label{line:rq3_orig_pre_end}@ 
p->next = &l.tail;
l.tail.prev = p;  
list_front(l);@\label{line:rq3_orig_cuv}@ 
Node *nnode = l.head.next;@\label{line:rq3_orig_post_start}@
for (int idx=0; idx < MAX; idx++) {@\label{line:rq3_orig_isvalid}@  
  nnode = nnode->next;  }
assert(nnode == l.tail);@\label{line:rq3_orig_post_end}@ 
\end{lstlisting}
\end{lrbox}

\newsavebox{\figelevenbox}
\begin{lrbox}{\figelevenbox}\begin{lstlisting}[escapechar=@]
linked_list l;
Node *n = malloc(sizeof(Node));@\label{line:rq3_new_node}@ 
n->next = nd_voidp();@\label{line:rq3_new_ndvoidp}@ 
l.head.next = n;
l.tail.prev = nd_voidp();
list_front(l);
assert(l.head.next == n);@\label{line:rq3_new_isvalid}@
\end{lstlisting}
\end{lrbox}

\begin{figure}[t]
\centering
\begin{subfigure}[t]{0.4\textwidth}
\scalebox{0.8}{\usebox{\figtenbox}}
\caption{Spec in the style of \cite{DBLP:conf/icse/ChongCKKMSTTT20}} 
\label{fig:rq3_orig}
\end{subfigure}
\hfill
\begin{subfigure}[t]{0.4\textwidth}
\scalebox{0.8}{\usebox{\figelevenbox}}
\caption{New specification} 
\label{fig:rq3_new}
\end{subfigure}
\caption{Simplified code for differing \CaS specifications.} 
\label{fig:rq3_orig_new}
\end{figure}

\newsavebox{\figeltwelvebox}
\begin{lrbox}{\figeltwelvebox}\begin{lstlisting}[escapechar=@]
  char buf[SZ];
  init_buf(buf, SZ);
  int idx = nd_int();@\label{line:rq3_buf_init_start}@
  assume(0 <= idx && idx < SZ);
  char saved = buf[idx];@\label{line:rq3_buf_init_end}@
  read_only_op(buf);@\label{line:rq3_read_only}@
  assert(saved == buf[idx]);@\label{line:rq3_buf_assert}@
\end{lstlisting}
\end{lrbox}

\newsavebox{\figelthirteenbox}
\begin{lrbox}{\figelthirteenbox}\begin{lstlisting}[escapechar=@]
  char buf[SZ];
  init_buf(buf, SZ);
  tracking_on();
  read_only_op(buf);@\label{line:rq3_is_mod_read_only}@
  assert(!is_mod(buf));
\end{lstlisting}
\end{lrbox}

\begin{figure}[t]
\centering
\begin{subfigure}[t]{0.4\textwidth}
\scalebox{0.8}{\usebox{\figeltwelvebox}}
\caption{Spec in style of~\cite{DBLP:conf/icse/ChongCKKMSTTT20}}
\label{fig:rq3_save_byte}
\end{subfigure}
\hfill
\begin{subfigure}[t]{0.4\textwidth}
\scalebox{0.8}{\usebox{\figelthirteenbox}}
\caption{\mbox{Spec using a built-in \code{is\_mod}}}
\label{fig:rq3_is_modified}
\end{subfigure}
\caption{Two styles of specifications for a read only buffer operation.} 
\label{fig:rq3_builtin}
\end{figure}
There are many alternative ways to express a specification in \CaS. 
In this section, we illustrate how to make proofs more efficient and make specs more readable.
For example, a unit proof can fully instantiate a data structure (as in a unit test), or minimally constrain it (as in~\cite{DBLP:conf/icse/ChongCKKMSTTT20}). In this section, we illustrate this by describing our experience in making \code{linked_list} unit proofs unbounded (and more efficient). Furthermore, we believe that extending the specification language with additional verifier-supported built-in functions simplifies specs while making them easier to verify. We illustrate this with the built-ins developed for \seahorn to specify absence of side-effects.

\paragraph{Linked List.}
A common pattern in \emph{unit proofs} is to assume the representation invariant 
of a data structure, and to assert it after invocation of the function under 
verification along with other properties that must be maintained by the function. 
For example, a simplified version of its unit proof from~\cite{DBLP:conf/icse/ChongCKKMSTTT20} is shown in \cref{fig:rq3_orig}.
The pre-conditions are specified by (explicitly) creating a list in lines~\ref{line:rq3_orig_create}--\ref{line:rq3_orig_pre_end}  using a loop. The post-condition is checked by completely traversing the list in lines~\ref{line:rq3_orig_isvalid}--\ref{line:rq3_orig_post_end}. This specification is simple since it closely follows the style of unit tests. However, it is inefficient for BMC: (a) unrolling the loops in the pre- and post-conditions blows up the symbolic search space; (b) it makes verification of the loop-free function \code{list_front} bounded, i.e., verification appears to depend on the size of the list in the pre-conditions.

\begin{figure}[t]
  \centering
  \includegraphics[scale=0.8]{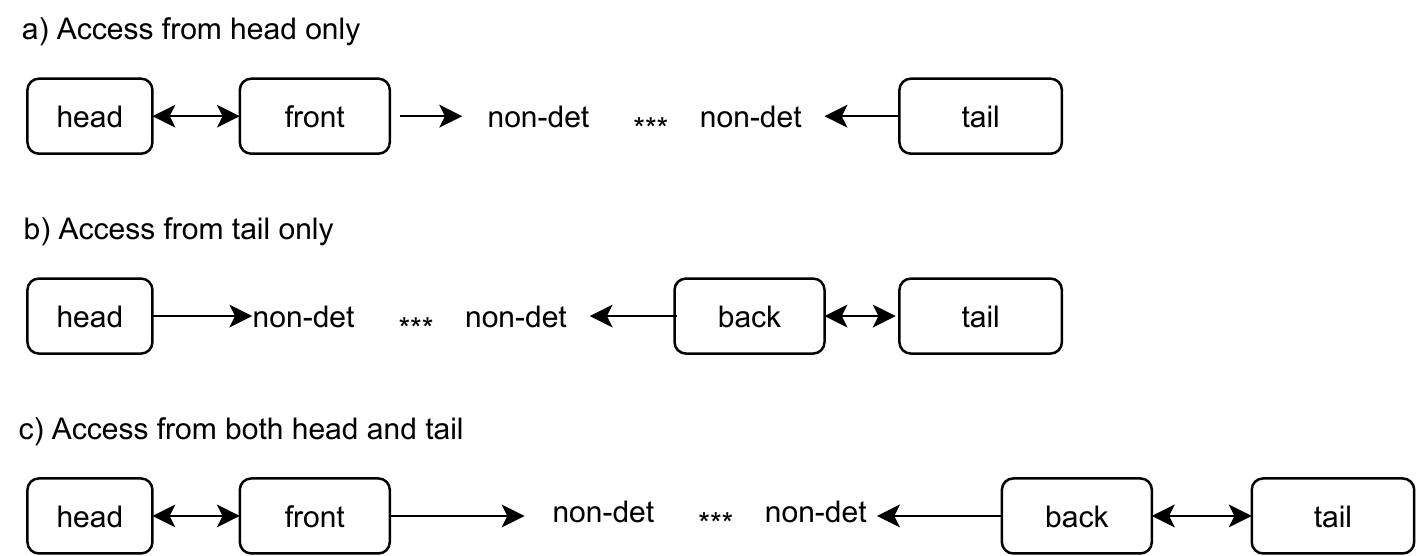}
  \caption{Linked list stubs for proofs.}
  \label{fig:dll}
\end{figure}

Our alternative formulation is to construct a partially defined linked list stub as shown in \cref{fig:dll}a. This stub can be used to verify \code{list_front} since it is expected that only the first node after head is accessed. The resulting \CaS is shown in \cref{fig:rq3_new}. The \code{next} field of \code{n} points to a potentially invalid address (returned by \code{nd_voidp}). Either \code{list_front} never touches \code{n->next} or has a memory fault. Finally, the assert on line~\ref{line:rq3_new_isvalid} in \cref{fig:rq3_new} checks that \code{list_front} did not modify the head of the list either. If there is no memory fault, then \code{list_front} did not modify the linked list after the node \code{n}. Our specification is not inductive. It uses the insight that the given linked list API only ever accesses a single element. It, therefore, avoids loops in both the pre- and post-conditions and verifies \code{list_front} for linked lists of any size.

Unfortunately, our new spec in \cref{fig:rq3_new} is difficult to understand by non-experts because it relies on the interplay between \code{nd_voidp} and memory safety checking.
To make the spec accessible, we hide the details behind a helper API. \cref{fig:linked_list_front_after} 
shows the unit proof for \code{aws_linked_list_front} with this API. 
The function \code{sea_nd_init_aws_linked_list_from_head} constructs partial \code{aws_linked_list} 
instances with non-deterministic length (as shown in \cref{fig:dll}a). The function \code{aws_linked_list_save_to_tail} 
saves concrete linked list nodes from the partial \code{aws_linked_list}. Finally, the function \code{is_aws_list_unchanged_to_tail} 
is used in post-conditions to check that linked list nodes are not modified. The unit proof for \code{aws_linked_list_front} is not only more efficient  than the original \cbmc proof, but it is also a 
\emph{stronger} specification. For example, if \code{aws_linked_list_front} 
removes or modifies a linked list node, our unit proof catches this as a violation, while the original proof only checks whether the returned value is valid and whether 
the linked list is well formed. The API we devised is generalized to work with all linked list operations in \code{aws-c-common}. For operations which access the node before the tail we construct a partially defined stub as shown in \cref{fig:dll}b while \cref{fig:dll}c is constructed for operations which access the list from both ends. We provide corresponding versions of the above API to save and check immutability of linked list nodes for each kind of stub.

\newsavebox{\figlistmainbox}
\begin{lrbox}{\figlistmainbox}\begin{lstlisting}
    void aws_linked_list_front_harness() {
      /* data structure */
      struct aws_linked_list list;
      struct saved_aws_linked_list to_save = {0};
      size_t size;

      sea_nd_init_aws_linked_list_from_head(&list, &size);
      struct aws_linked_list_node *start = &list.head;
      aws_linked_list_save_to_tail(&list, size, start, &to_save);

      // precondition in function does not accept empty linked list
      assume(!aws_linked_list_empty(&list));

      /* perform operation under verification */
      struct aws_linked_list_node *front = aws_linked_list_front(&list);

      /* assertions */
      sassert(list.head.next == front);
      sassert(aws_linked_list_node_prev_is_valid(front));
      sassert(aws_linked_list_node_next_is_valid(front));
      sassert(is_aws_list_unchanged_to_tail(&list, &to_save));

      return 0;
    }
  \end{lstlisting}
\end{lrbox}

\begin{figure}
  \centering
  \scalebox{0.8}{\usebox{\figlistmainbox}}
  \caption{\seahorn unit proof for \code{aws_linked_list_front}.} 
  \label{fig:linked_list_front_after}
\end{figure}

\paragraph{Increasing \CaS expressiveness.} Verification tools should provide built-ins to aid in concise specifications. Moreover, such built-ins enable specifications that are not otherwise expressible in \CaS. 
For example, \cref{fig:rq3_is_modified} uses a \seahorn built-in, \code{is_mod}, to specify that \code{read_only_op} does not change the buffer.
This built-in returns true if memory pointed by its argument is modified since the last call to \code{tracking_on}.
In contrast, the original specification for \cbmc in \cref{fig:rq3_save_byte} is tricky. 
It saves a byte from some position in the buffer (lines~\ref{line:rq3_buf_init_start}--\ref{line:rq3_buf_init_end}), and checks that it is not changed (line~\ref{line:rq3_buf_assert}). This example illustrates that built-ins make specifications simpler and more direct.
They ease specification writing for users and might be exploited efficiently by verification tools.
As another example, \seahorn provides a built-in \code{is_deref} to check that a memory access is within bounds, which is not (easily) expressible in C.

\paragraph{Discussion.} \CaS enables concise specifications and efficient proofs.
As advanced verification techniques may not generalize, a standard extension is needed, such as verification-specific built-in functions. The semantics of these can be provided by a run-time library, validated by fuzzing and supported by multiple verification tools. Additional case studies are needed to identify a good set of built-ins. A standard extension can increase specification reuse and make verification more productive and effective.

  \section{Related work}
\label{sec:related}

To our knowledge,~\cite{DBLP:conf/icse/ChongCKKMSTTT20} is the first significant, publicly available, example of an application of BMC on industrial code that is actively maintained with the code. Thus, our work is the first exploration of potential issues with software verified in this way. 
The closest verification case studies are \texttt{coreutils} with \klee~\cite{DBLP:conf/osdi/CadarDE08} and \texttt{busybox} \texttt{ls} with CBMC~\cite{DBLP:conf/apsec/KimK14}. However, those focus on the scalability of a specific verification technology, while we focus on methodology, reuse, and what bugs might be hidden in the verification effort. 

As we mentioned in the introduction, the Software Verification Competition (SV-COMP)~\cite{DBLP:conf/tacas/Beyer20} provides a large collection of benchmarks, and, an annual evaluation of many verification tools. However, it is focused on performance and soundness of the tools. The benchmarks are pre-processed to fit the competition format. At that stage, it is impossible to identify and evaluate the specifications, or to modify the benchmarks to increase efficiency of any particular tool. We hope that our case study can serve as an alternative benchmark to evaluate suitability of verification tools for industrial transition.

In addition to~\cite{DBLP:conf/icse/ChongCKKMSTTT20}, there has been number of other recent applications of BMC at AWS, including~\cite{DBLP:conf/cav/CookKKTTT18,DBLP:conf/cav/ChudnovCCDHMMMM18,DBLP:conf/fmcad/CookDKMPPTW20}. However, they are either not publicly available, too specialized, or not as extensive as the case study in~\cite{DBLP:conf/icse/ChongCKKMSTTT20}.

Using code as specification has a long history in verification tools, one prominent example is Code Contracts introduced in Spec$^{\sharp}$~\cite{DBLP:journals/cacm/BarnettFLMSV11}. One important difference is that in our case \CaS is used to share specifications between completely different tools that only share the semantics of the underlying programming language, and the language itself is used to adapt specifications to the tools.

 \vspace{-0.1in}
\section{Conclusion}
\vspace{-0.1in}
\label{sec:conclusion}
This case study would not have been possible without artifacts released by AWS in \cite{DBLP:conf/icse/ChongCKKMSTTT20}. To our knowledge, it is the first publicly available application of BMC (to software) in industry. Related case studies on verification are those on \texttt{coreutils} with \klee~\cite{DBLP:conf/osdi/CadarDE08} and on \texttt{busybox} \texttt{ls} with CBMC~\cite{DBLP:conf/apsec/KimK14}. SV-COMP is a large repository of benchmarks, but its goals are different from an actively maintained industrial project. 
The availability of both methodology and artifacts has given us a unique opportunity to study how verification is applied in industry and to improve verification methodology.  We encourage industry to release more benchmarks to enable further studies by the research community.

In addition to answering the research questions, we are contributing a complete working system that might be of interest to other researchers. We have implemented a custom build system using \textsc{CMake} that simplifies integrating new tools. We provide Docker containers to reproduce all of the results. We created continuous integration (CI) on GitHub that nightly re-runs all the tools on the current version of \awsccommon. Since we use standard tools, the project integrates seamlessly into IDEs and refactoring tools. The CI runs are done in parallel by \textsc{CTest}. Running \seahorn takes under $8$ minutes!

While comparing different tools on performance is not our primary concern, in
\cref{tab:time-comp}, we show the running time for all of the verification
tools, collected on the same machine. \ifshort\else Runtime for individual unit
proofs are shown in \cref{sec:results} (and online). \fi For \libfuzzer, we make the detailed coverage report available online. We stress that while the tools check the same explicit assertions, they check different built-in properties. Thus, running time comparison must be taken with a grain of salt.

Our main conclusion is in agreement with \cite{DBLP:conf/icse/ChongCKKMSTTT20},  and we strengthen the evidence for it. \CaS is a practical and scalable approach for specifications that is easy to understand and empowers many tools.  We argue that using de facto compiler semantics in \CaS is key for enabling many verification tools, each with its own characteristic, to be used on the same verification problem. We find that specifications can be written in different ways and  specification writer must account for both verification efficiency and developer readability. We suggest that a set of common built-ins be shared by different verification tools. Such built-ins improve the expressive power of \CaS while \mbox{retaining portability across verification tools.}
 With built-ins defined in a specification library, software developers will be able to write unit proofs in a way no difference than programming with libraries provided by some framework.

Today, formal verification is not the primary means of building confidence in software quality. Our hope is that case studies like this one are useful to both software engineering researchers and practitioners who want to make formal methods an integral part of software development. To further this agenda, we plan to continue applying the \CaS methodology to larger and more complex code bases (and languages) in the future.

 \bibliographystyle{splncs04}

\ifshort
\else
\newpage
\appendix
\section{Description of Specification Bugs Found}
\label{sec:bugs}
In this section, we describe the bugs that we have found in the unit proofs of \awsccommon.
\subsection{Representation Invariant of \texttt{byte\_buf}}
\label{sec:bug-byte_buf}

\paragraph{File. } \code{source/byte_buf.c}

\paragraph{Description.} The bug is in the representation invariant of the
\code{byte_buf} data structure. The code is shown in
\cref{fig:byte_buf_is_valid}, and the revision is shown in the comment. The
\code{byte_buf} data structure contains a field \code{buffer} that points to a
writable memory region of size at least \code{capacity}. In the representation
invariant, \code{len} was used instead of \code{capacity}. This allows for
\code{buffer} to be \code{NULL} when \code{len} is $0$ and \code{capacity} is
not zero. The buggy specification is \emph{not} an invariant. We showed that it is
not maintained by the API of \code{byte_buf}.

The representation invariant is used both as pre- and post-condition in unit
proofs, and also in unit tests. The buggy version is weaker than the correct
one. Thus, it is true of any correct instance of \code{byte_buf}. For this
reason, it is the bug that has not been detected by unit tests.

\paragraph{Story.} According to GitHub logs, the bug was in the code for at
least two years. We found it indirectly using vacuity in \seahorn. We
have initially modelled memory allocation (i.e., calls to \code{malloc}) as
never failing. That is, never returning a \code{NULL} pointer. Under this
assumption, some post-conditions became vacuous. We have then re-defined
\code{malloc} to non-deterministically fail. This caused one of the unit proofs
that was using this representation invariant to fail with a counterexample.
Examining the counterexample manually lead to the discovery of the bug. 

The bug was reported to AWS. It has pointed to a deeper issue of how
\code{malloc} should be modelled. This resulted in review of all CBMC-based
proofs at AWS with numerous issues uncovered and fixed. The bug is fixed in
\awsccommon in commit
\texttt{ec70687}\footnote{\url{https://github.com/awslabs/aws-c-common/pull/686}.}.

\newsavebox{\figbytebufbox}
\begin{lrbox}{\figbytebufbox}\begin{lstlisting}
bool aws_byte_buf_is_valid(const struct aws_byte_buf *const buf) {
  return buf != NULL &&
  ((buf->capacity == 0 && buf->len == 0 && buf->buffer == NULL) ||
  (buf->capacity > 0 && buf->len <= buf->capacity &&
  AWS_MEM_IS_WRITABLE(buf->buffer, buf->len /* buf->capacity */)));
}
\end{lstlisting}
\end{lrbox}

\begin{figure}[t]
\centering
\scalebox{0.8}{\usebox{\figbytebufbox}}  
\caption{Representation invariant of \code{aws_byte_buf}.} 
\label{fig:byte_buf_is_valid}
\end{figure}

 \subsection{Bug in verification helper \code{assert_bytes_match}}
\label{sec:bug-assert_bytes_match}

\paragraph{File. }\code{verification/cbmc/sources/utils.c}

\paragraph{Description.} The bug in verification library function that checks that two
byte regions are equivalent. The bug and the revision (in comments) are shown in
\cref{fig:assert_bytes_match}. This function is used to compare content of
zero-terminated strings with non-zero terminated byte buffers. A zero sized
string is not \code{NULL} (it needs one character for the zero-terminator). A
zero sized byte buffer is \code{NULL} because \awsccommon ensures that
allocating 0 bytes returns \code{NULL}. Thus, the buggy version of the function
returns false when comparing an empty string with an empty byte buffer. This
contradicts the name and the use of the function in unit proofs. 

\paragraph{Story.} According to GitHub logs, the bug was in the code for at
least two years. We found the bug by allowing \code{malloc} to fail and return
\code{NULL} non-deterministically. This enabled code paths in existing unit
proofs are infeasible under previous assumptions. This caused some unit
proofs to fail. Examining the counterexamples manually lead to discovery of this
bug.

\newsavebox{\figbytematchbox}
\begin{lrbox}{\figbytematchbox}\begin{lstlisting}
void assert_bytes_match(const uint8_t *const a, 
                        const uint8_t *const b, const size_t len) {
  assert(/* len == 0  || */ !a == !b); 
  if (len > 0 && a != NULL && b != NULL) {
    size_t i;
    /* prevent spurious pointer overflows */
    __CPROVER_assume(i < len && len < MAX_MALLOC); 
    assert(a[i] == b[i]);
  }
}
  \end{lstlisting}
\end{lrbox}

\begin{figure}[t]
  \centering
  \scalebox{0.8}{\usebox{\figbytematchbox}}
  \caption{Verification library function \code{assert_bytes_match}.} 
  \label{fig:assert_bytes_match}
\end{figure}

 \subsection{Post-condition bug in unit proof for \code{aws_mul_size_checked_harness.c}}

\paragraph{File.} \code{verification/cbmc/proofs/aws_mul_size_checked/}\\\code{aws_mul_size_checked_harness.c}

\paragraph{Description.} The bug is in the post-condition of the unit proof. The bug and the revision (in comments) are shown in
\cref{fig:aws_mul_size_checked}. The function \code{aws_mul_size_checked} is a checked overflow multiplication. If the multiplication does not
overflow, the result of the multiplication is stored in \code{r}, otherwise, the
function returns an error code. The assertion at line~16 is intended to check
for overflow, but instead, it checks for overflow of addition. The verification
environment is restricted at line~9 to two special values. The
post-condition happens to be true for these two inputs.

\paragraph{Story.} The comments in the unit proof restrict the environment because of scalability issues. This was not a problem for
\seahorn. When migrating, we have removed the assumption at line~9 and found a
counterexample. 

The bug was fixed by AWS team as shown in the commented assertion at line~16.
The revision, arguably, creates a different bug. When the code if \awsccommon is
compiled for CBMC, it redefines \code{aws_mul_u64_checked} with a call to
\code{__CPROVER_overflow_mult}. Thus, this unit proof does not check any actual
executable code in the library.

The library provides several implementations of checked arithmetic. The most
common one uses checked arithmetic builtins provided by both \textsc{Clang} and
\textsc{GCC} compilers, and others rely on inline assembly. In \seahorn, we
verify the version that is compiled using compiler builtins. Thus, with a
similar unit proof, \seahorn is able to check correctness of the actual
implemented code.

\newsavebox{\figmulbox}
\begin{lrbox}{\figmulbox}\begin{lstlisting}
#include <aws/common/math.h>
extern int nondet_int(void);
extern uint64_t nondet_uint64_t(void);
void aws_mul_size_checked_harness() {
  /*
  * In this particular case, full range of nondet inputs leads
  * to excessively long runtimes, so use 0 or UINT64_MAX instead.
  */
  uint64_t a = (nondet_int()) ? 0 : UINT64_MAX;
  uint64_t b = nondet_uint64_t();
  uint64_t r = nondet_uint64_t();
  if (!aws_mul_u64_checked(a, b, &r)) {
    assert(r == a * b);
  } else {
    assert((b > 0) && (a > (UINT64_MAX - b)));
    /*  assert(__CPROVER_overflow_mult(a, b));
  }
}
  \end{lstlisting}
\end{lrbox}

\begin{figure}[t]
  \centering
  \scalebox{0.8}{\usebox{\figmulbox}}
  \caption{Unit proof for \code{aws_mul_size_checked}.} 
  \label{fig:aws_mul_size_checked}
\end{figure}

 \subsection{Post-condition bug in unit proof for \code{s_swap} of priority queue}
\label{sec:pq_s_swap}

\paragraph{File.}\code{verification/cbmc/proofs/aws_priority_queue_s_swap/}\\\code{aws_priority_queue_s_swap_harness.c}

\paragraph{Description.} The bug is in one of the post-conditions of the
function \code{s_swap} that is used in the implementation of the priority queue data
structure. The complete unit proof is too long to reproduce here, so we present a
relevant snippet in \cref{fig:s_swap} instead. Since the function \code{s_swap}
is private to the priority queue translation unit, a special syntax is used to
call it directly from the unit proof (line~4). The bug is in the condition of
the if-statement at line~8. Conjunction is used instead of disjunction. The revision
is shown in comments and corresponds to the comment at line~6.

\paragraph{Story.} The bug was found by vacuity detection. The assertions inside the function
\code{assert_array_list_equivalence} are unreachable. The bug has been reported,
but is not yet fixed. GitHub logs show that the bug has been active for at least
two years. The coverage results from CBMC show that the call to
\code{assert_array_list_equvalence} is not covered (i.e., not executed by the
unit proof). However, this seems to have been missed. 

\newsavebox{\figswapbox}
\begin{lrbox}{\figswapbox}\begin{lstlisting}
void aws_priority_queue_s_swap_harness() {
  ...
  /* Perform operation under verification */
  __CPROVER_file_local_priority_queue_c_s_swap(&queue, a, b);
  ...
  /* All the elements in the container except for a and b should stay 
     unchanged */
  size_t ob_i = old_byte.index;
  if ((ob_i < a * item_sz && /* || */ ob_i >= (a + 1) * item_sz) &&
      (ob_i < b * item_sz && /* || */ ob_i >= (b + 1) * item_sz)) {
    assert_array_list_equivalence(&queue.container, &old, &old_byte);
  }
  ...
}
    \end{lstlisting}
\end{lrbox}

\begin{figure}[t]
  \centering
  \scalebox{0.8}{\usebox{\figswapbox}}
  \caption{Snippet of the unit proof for \code{s_swap} of priority queue.} 
  \label{fig:s_swap}
\end{figure}
 \subsection{Pre-condition bug mixing different kinds of strings}

\paragraph{File.}\code{verification/cbmc/proofs/aws_hash_callback_string_eq/}\\
\code{hash_callback_string_eq_harness.c}

\paragraph{Description.} The \awsccommon library supports two types of strings.
The zero-terminated strings from \textsc{LibC} that are common in C, and its own
data structures \code{aws_string}. The harness shown in
\cref{fig:aws_hash_callback_string} assumes that the \code{aws_string} object
satisfies the C-string representation invariant. The representation invariant
for C-strings happens to be very weak (essentially that the pointer is not
null). According to the C semantics, conversion from a data structure to a
C-string is always possible. 
\paragraph{Story.} The bug was found during compilation in an IDE. Compilation
warnings indicated that an explicit cast is desirable for the calls to
\code{aws_c_string_is_valid}. It is clear that the intended specification is to
assume representation invariant for \code{aws_string} instead.

\newsavebox{\figcallbackbox}
\begin{lrbox}{\figcallbackbox}\begin{lstlisting}
void aws_hash_callback_string_eq_harness() {
  const struct aws_string *str1 = ...; 
  const struct aws_string *str2 = ...; 

  __CPROVER_assume(aws_c_string_is_valid(str1));
  __CPROVER_assume(aws_c_string_is_valid(str2));
  /* __CPROVER_assume(aws_string_is_valid(str1)); */
  /* __CPROVER_assume(aws_string_is_valid(str2)); */



  bool rval = aws_hash_callback_string_eq(str1, str2);
  if (rval) {
    assert(str1->len == str2->len);
    assert_bytes_match(str1->bytes, str2->bytes, str1->len);
  }
}
\end{lstlisting}
\end{lrbox}

\begin{figure}[t]
\centering
\scalebox{0.8}{\usebox{\figcallbackbox}}  
\caption{Snippet of the unit proof for \code{aws_hash_callback_string_eq}.} 
\label{fig:aws_hash_callback_string}
\end{figure}

 \subsection{Specification stub of \texttt{hash\_iter\_delete}}
\label{sec:bug-hash_iter}

\paragraph{File.} \code{verification/cbmc/stubs/aws_hash_iter_overrides.c}

\paragraph{Description.} A bug in \emph{specification stub} that we have found in this case study is in the stubbed version of \code{hash_iter_delete}. The code is shown in \cref{fig:hash_iter_delete}, and the revision is shown in the comment at line 8. After deleting an item from the iterator, the specification stub did not update the \code{hash_table_state} of the hash table properly since the stub did not decrease the \code{entry_count} field of it. 

\paragraph{Story.} We found the bug in an attempt to use original functions over specification stubs in order to build a more precise proof of \code{hash_table_foreach}. The function \code{hash_table_foreach} leverages \code{hash_iter_begin}, \code{hash_iter_next} and \code{hash_iter_done} to iterate over all entries in an \code{aws_hash_table}. In each iteration, \code{hash_iter_delete} is conditionally invoked to delete an entry from the table. 

When we used the original \code{hash_iter*} functions in the \seahorn unit proof of \code{hash_iter_foreach}, we found that the representation invariant of \code{aws_hash_table} could become false in the post-condition of \code{hash_iter_foreach}. A deeper look at the counterexample trace revealed that the \code{entry_count} field became a negative number after a particular execution of \code{hash_iter_foreach}. 

The reason is two-fold. Firstly, there are two ways in \awsccommon hash table implementation to determine the number of entries in \code{aws_hash_table}. The first one is to explicitly look up the field \code{entry_count} and the second one is to implicitly infer it from the number of allocated entries that has a non-zero hash code. The original implementation of \code{hash_iter*} function often use the latter way. Secondly, the \cbmc proof non-deterministically initializes the memory that contains allocated hash table entry slots, but it does not set any constraint on the number of allocated entries with non-zero hash codes. 

When we tried to use original \awsccommon implementations of \code{hash_iter*} functions in combination of the initialization process from CBMC proofs, it is possible to create an \code{aws_hash_table} with more entries with non-zero hash codes than \code{entry_count}, in which case \code{hash_iter_delete} could be invoked more times than \code{entry_count} and decrease the field to a negative number.

The aforementioned behavior constitutes a bug in the pre-condition part of \emph{CaS}, which stems from a bug in a specification stub of key functions invoked by the function under verification.

\newsavebox{\fighashiterbox}
\begin{lrbox}{\fighashiterbox}\begin{lstlisting}
void aws_hash_iter_delete(struct aws_hash_iter *iter, 
                          bool destroy_contents) {
    /* Build a nondet iterator, set the required fields, 
       and copy it over */
    struct aws_hash_iter rval;
    rval.map = iter->map;
    rval.slot = iter->slot;
    rval.limit = iter->limit - 1;
    rval.status = AWS_HASH_ITER_STATUS_DELETE_CALLED;
    // rval.map->p_impl->entry_count--;
    *iter = rval;
}
\end{lstlisting}
\end{lrbox}

\begin{figure}[t]
\centering
\scalebox{0.8}{\usebox{\fighashiterbox}}  
\caption{Specification stub of \code{hash_iter_delete}.} 
\label{fig:hash_iter_delete}
\end{figure}
 \subsection{Undefined behaviour in \code{aws_is_mem_zeroed}}
\paragraph{File.} \code{include/aws/common/zero.inl}

\paragraph{Description.} A snippet of a buggy implementation of
\code{aws_is_mem_zeroed}  is shown in \cref{fig:is_mem_zeroed}. The function can
return false even when the memory pointed by \code{buf} is filled with zeroes.
The bug in the memory access in line~12. The buffer \code{buf} is accessed using
a pointer to \code{uint64_t}. This access is only defined if \code{buf} was
previously written with the same type. The code violates the strict aliasing
rule of C.  A suggested fix is given in comments. We do not know whether the bug
manifests in the production code. The TBAA alias analysis of LLVM is strong enough
to commute read at line~12 with prior writes to \code{buf}. 

\paragraph{Story.} In \awsccommon, the function is called through a macro
expansion of \code{AWS_IS_ZEROED}. The macro is used in unit proofs and in the
production code. In \cbmc mode, the macro expands to CBMC specific implementation.
Thus, the code of \code{aws_is_mem_zeroed} is never seen by CBMC. In \seahorn,
we analyze the code as it is compiled and this function is used. The bug in the
function causes a post-condition to be violated and counterexample is generated.
Manually examining the counterexample identified the bug.
\newsavebox{\figmemzerobox}
\begin{lrbox}{\figmemzerobox}\begin{lstlisting}
/**
 * Returns whether each byte is zero.
 */
AWS_STATIC_IMPL
bool aws_is_mem_zeroed(const void *buf, size_t bufsize) {
  const uint64_t *buf_u64 = (const uint64_t *)buf;
  const size_t num_u64_checks = bufsize / 8;
  size_t i;
  /* uint64_t val; */
  for (i = 0; i < num_u64_checks; ++i) {
    /* memcpy(&val, &buf_u64[i], sizeof(val)); */
    if (buf_u64[i] /* val */) {
      return false;
    }
  }
  ...
}
  \end{lstlisting}
\end{lrbox}

\begin{figure}[t]
  \centering
  \scalebox{0.8}{\usebox{\figmemzerobox}}
  \caption{A snippet of \code{aws_is_mem_zeroed}.} 
  \label{fig:is_mem_zeroed}
\end{figure}

\section{Empirical Results}
\label{sec:results}

\begin{figure}[t]
  \centering
\begin{subfigure}{.5\linewidth}
  \includegraphics[scale=0.38]{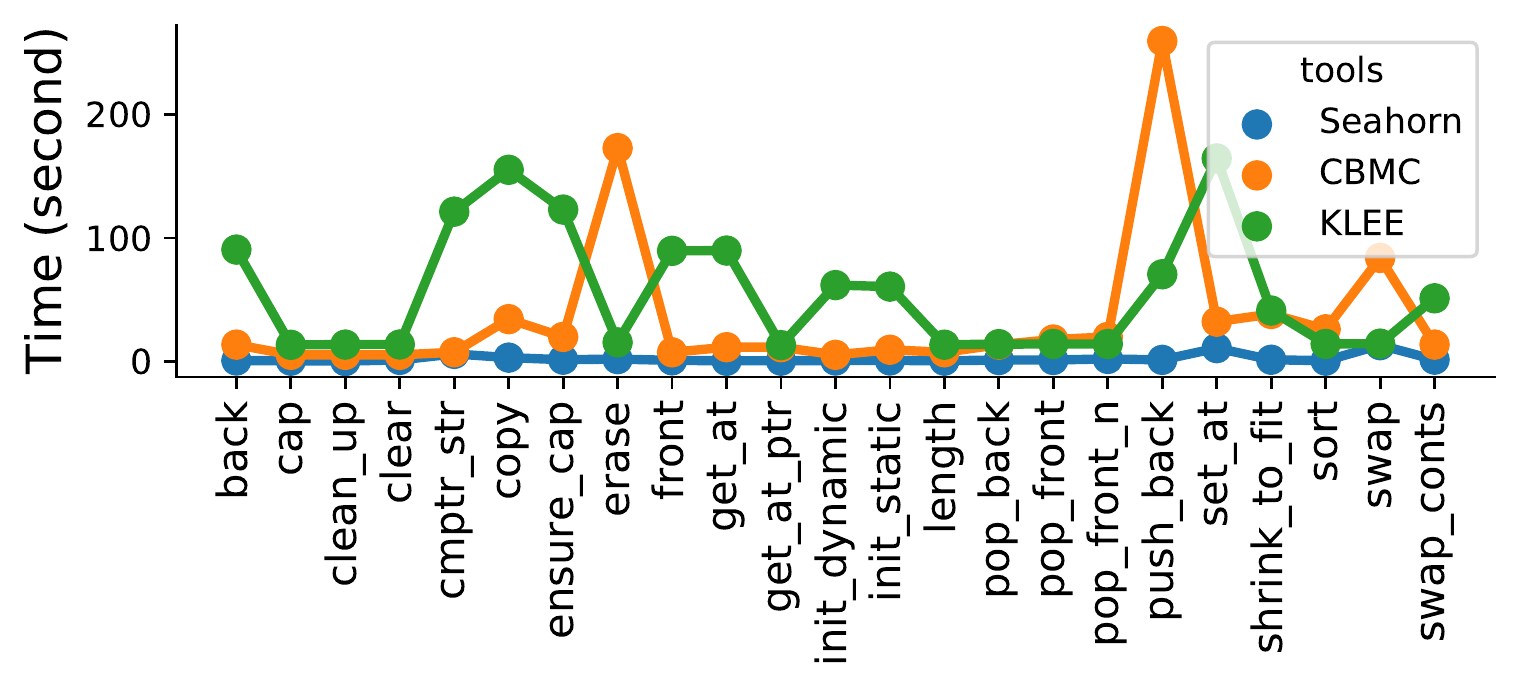}
  \caption{\code{aws_array_list}}
  \label{fig:comp_arylst}
\end{subfigure}\begin{subfigure}{.5\linewidth}
  \includegraphics[scale=0.35]{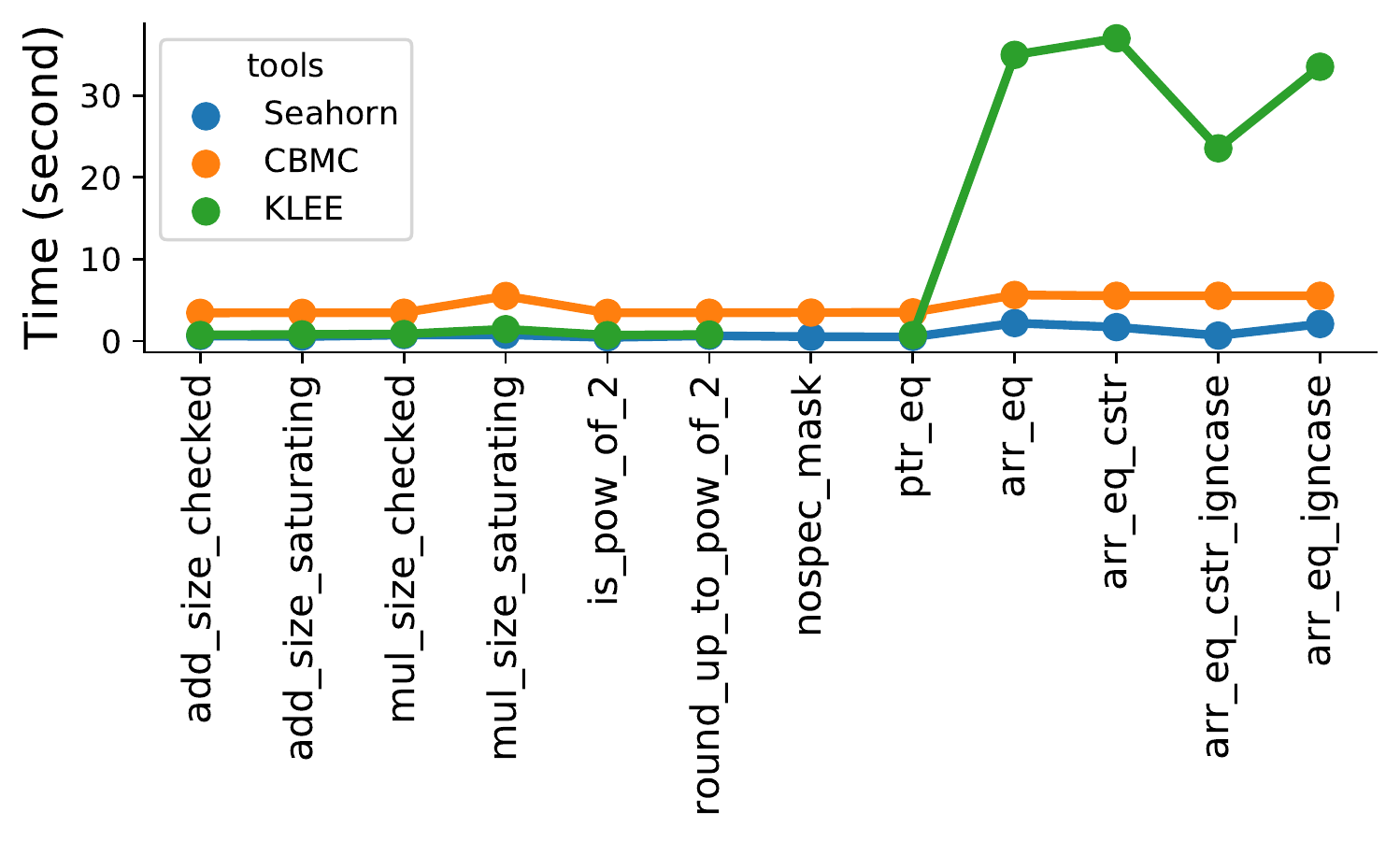}
  \caption{Others}
  \label{fig:comp_others}
\end{subfigure}\hfill
\begin{subfigure}{.5\linewidth}
  \includegraphics[scale=0.38]{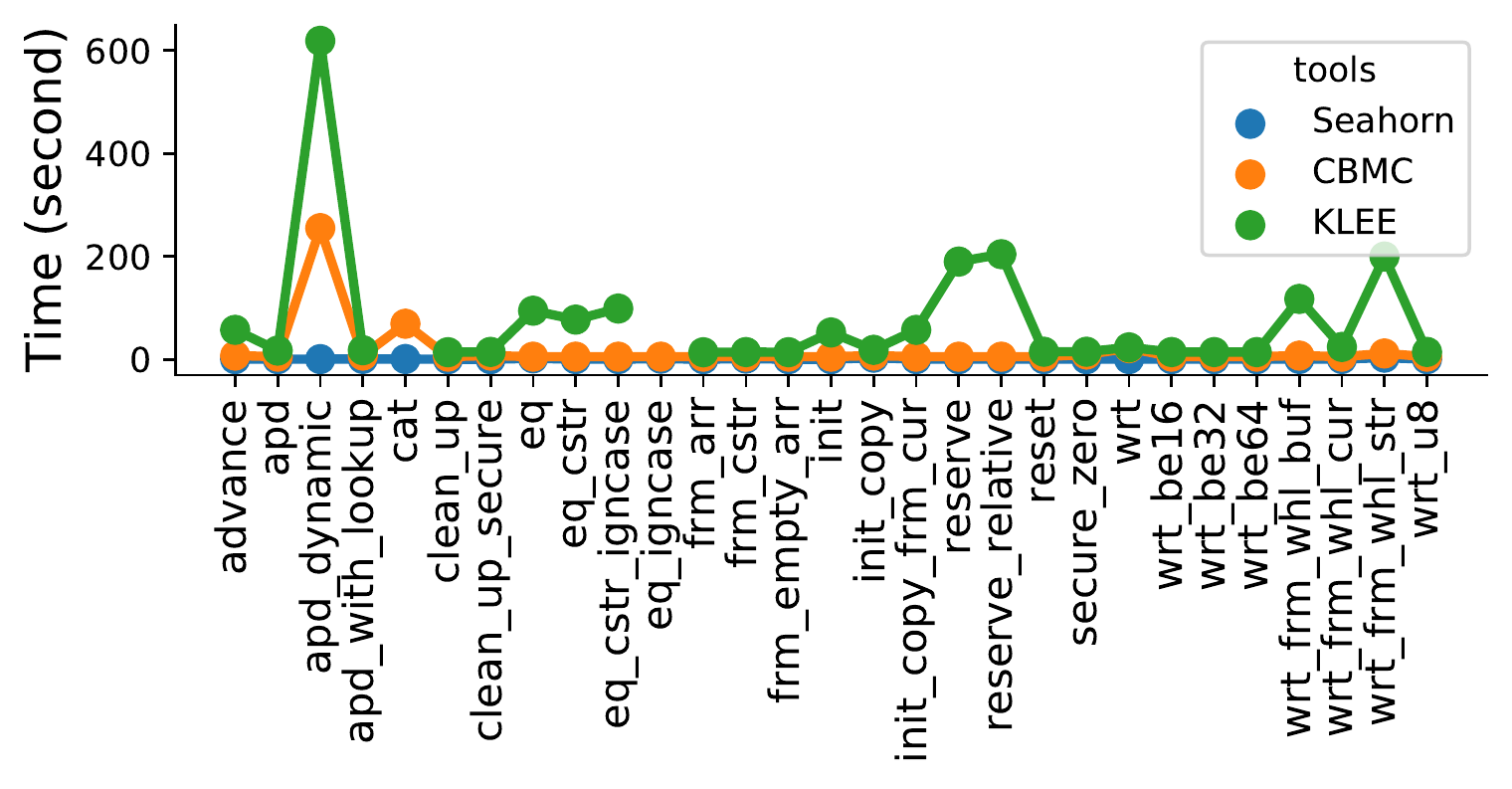}
  \caption{\code{aws_byte_buffer}}
  \label{fig:comp_bytebuf}
\end{subfigure}\begin{subfigure}{.5\linewidth}
  \includegraphics[scale=0.35]{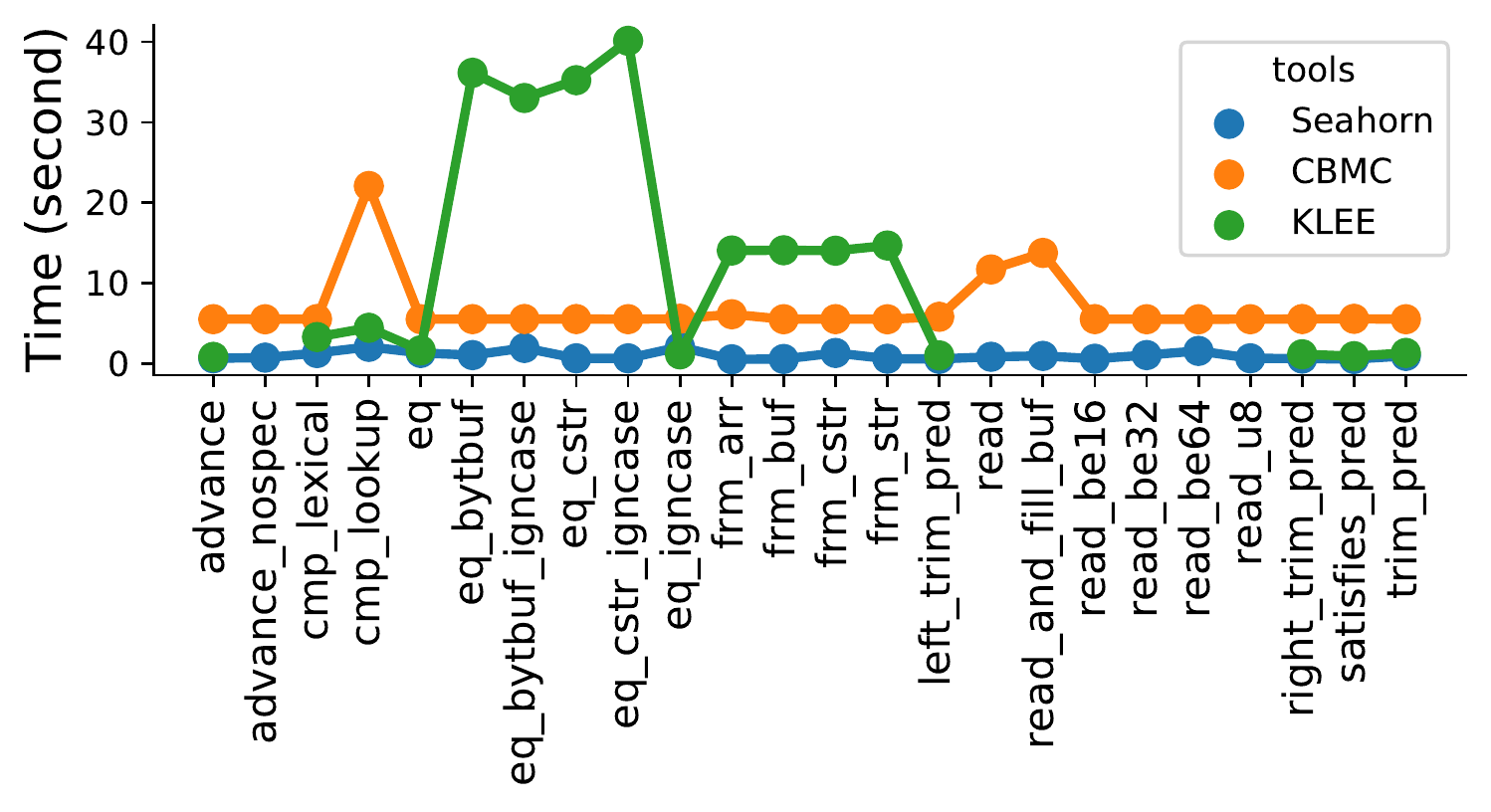}
  \caption{\code{aws_byte_cursor}}
  \label{fig:comp_bytecursor}
\end{subfigure}\hfill
\begin{subfigure}{.5\linewidth}
  \includegraphics[scale=0.38]{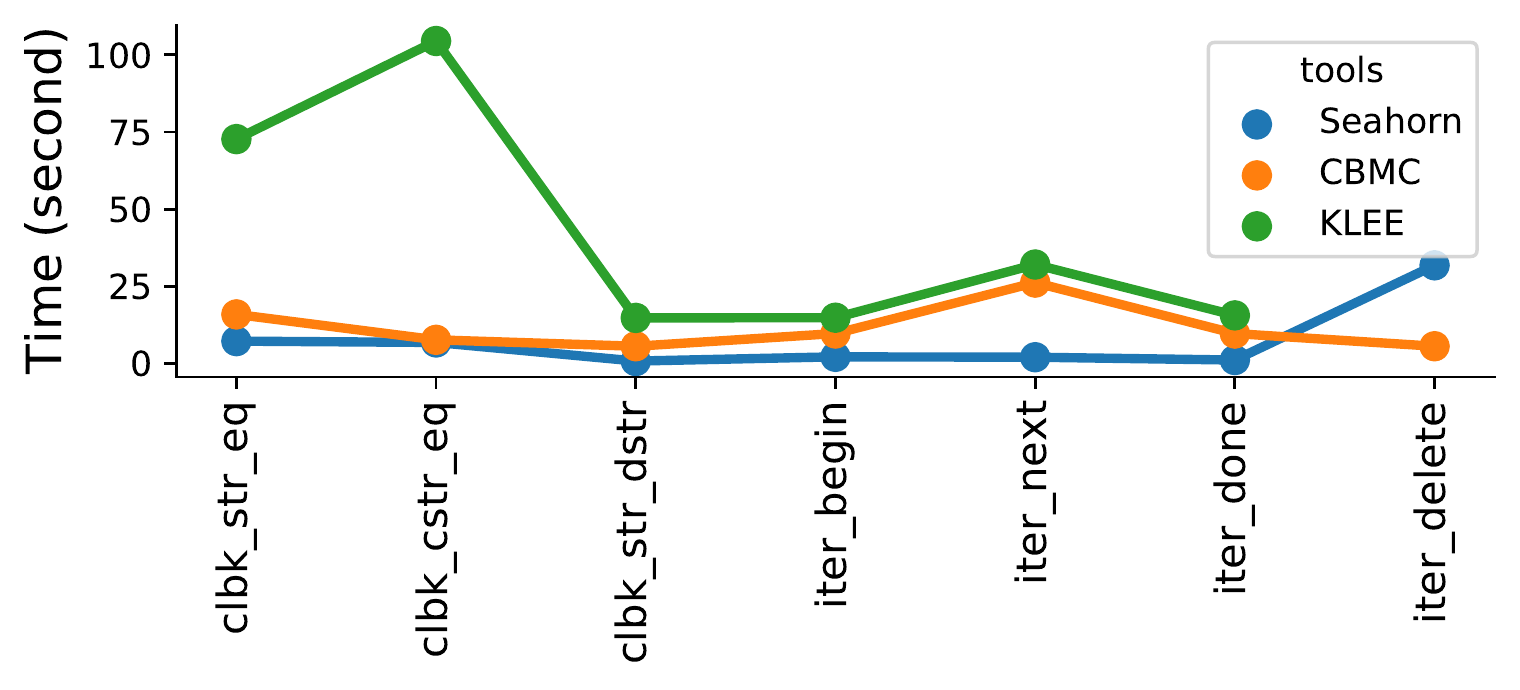}
  \caption{Hash}
  \label{fig:comp_hash}
\end{subfigure}\begin{subfigure}{.5\linewidth}
  \includegraphics[scale=0.35]{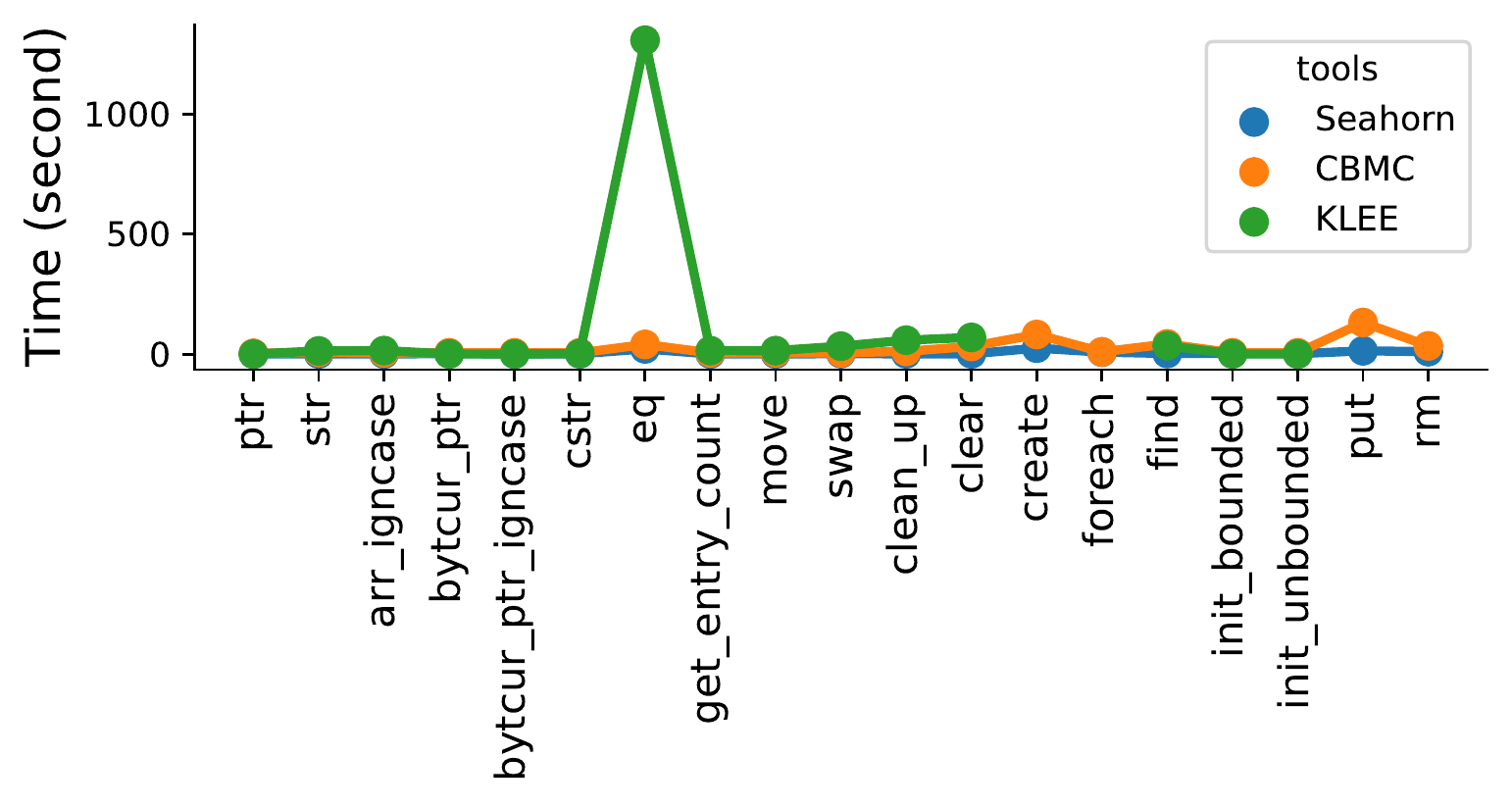}
  \caption{\code{aws_hash_table}}
  \label{fig:comp_hashtab}
\end{subfigure}\hfill
\begin{subfigure}{.5\linewidth}
  \includegraphics[scale=0.35]{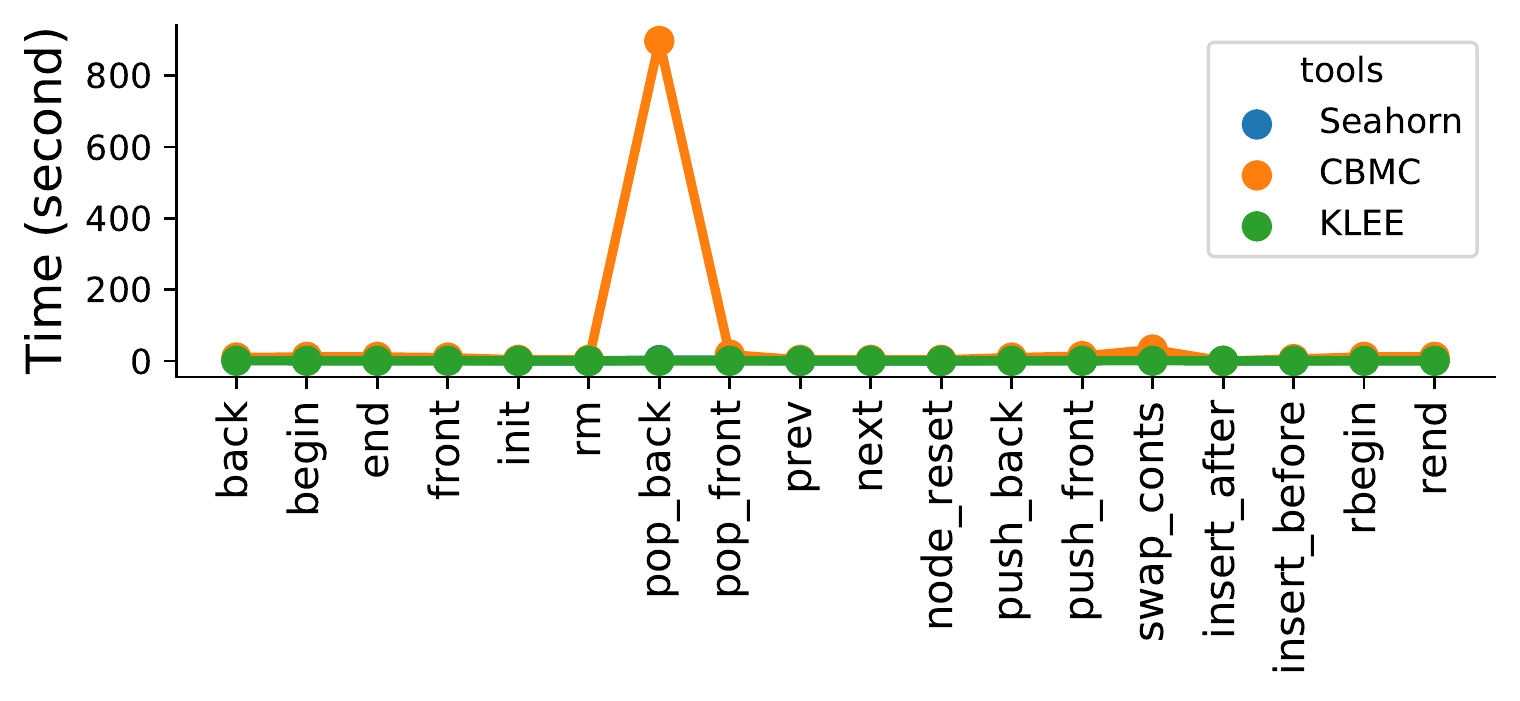}
  \caption{\code{aws_linked_list}}
  \label{fig:comp_linklst}
\end{subfigure}\begin{subfigure}{.5\linewidth}
  \includegraphics[scale=0.38]{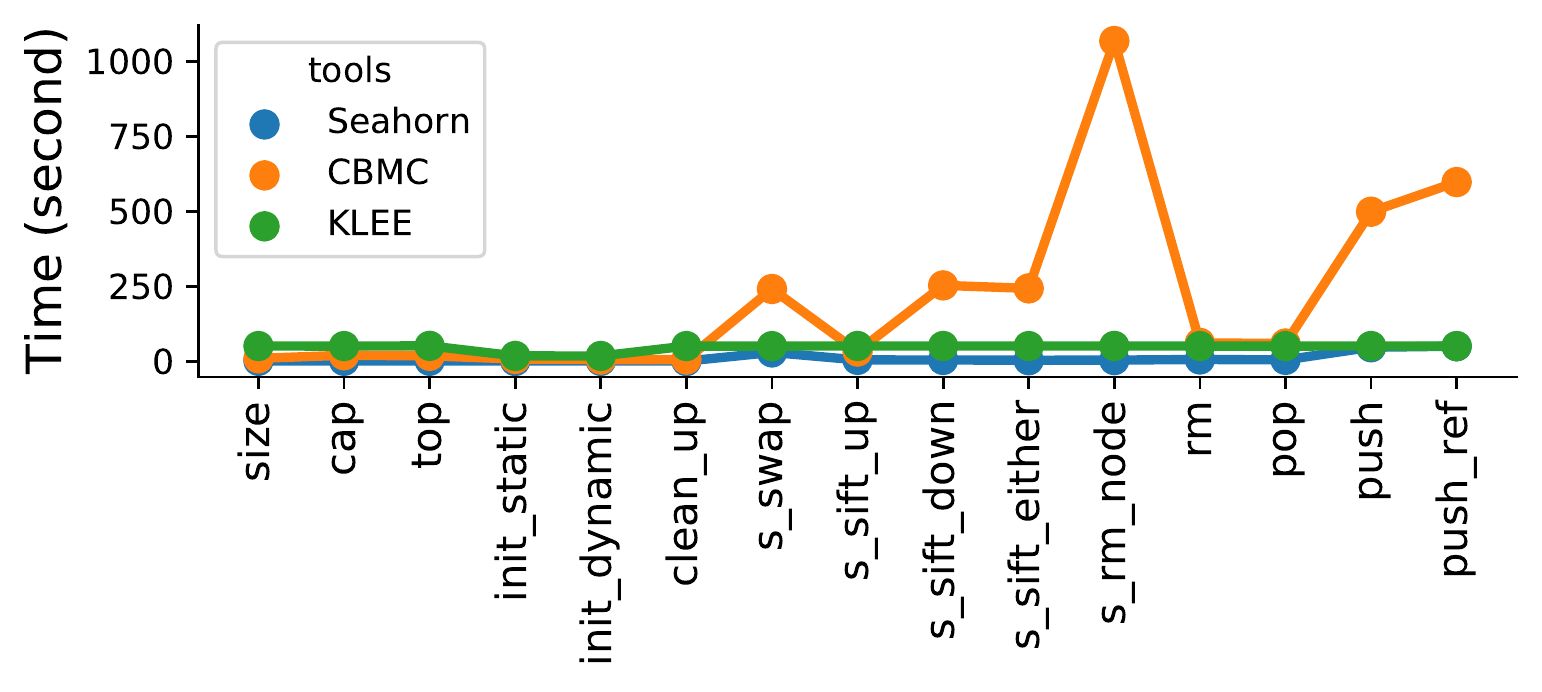}
  \caption{\code{aws_priority_queue}}
  \label{fig:comp_prioque}
\end{subfigure}\hfill
\begin{subfigure}{.5\linewidth}
  \includegraphics[scale=0.38]{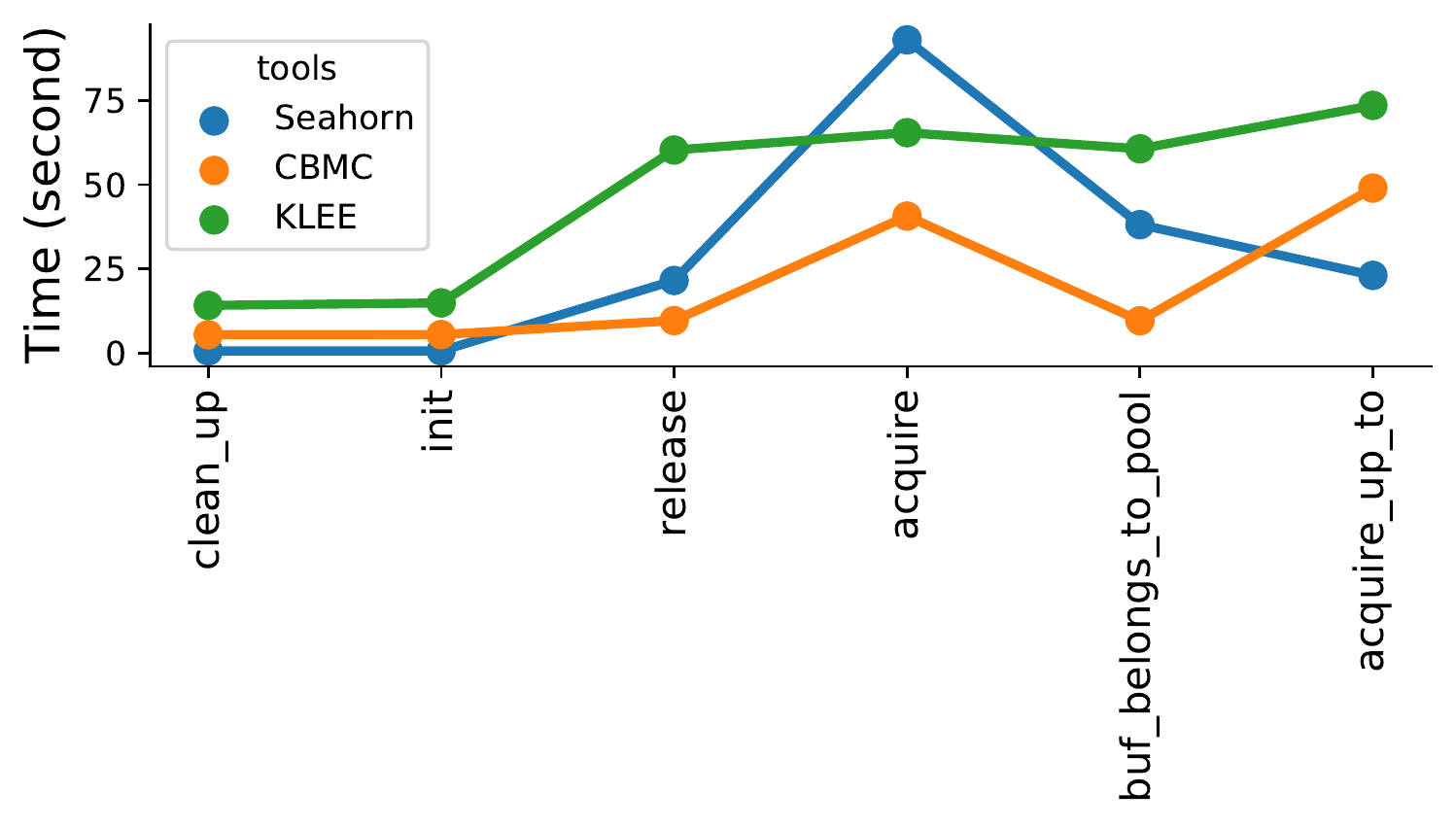}
  \caption{\code{aws_ring_buffer}}
  \label{fig:comp_ringbuf}
\end{subfigure}\begin{subfigure}{.5\linewidth}
  \includegraphics[scale=0.35]{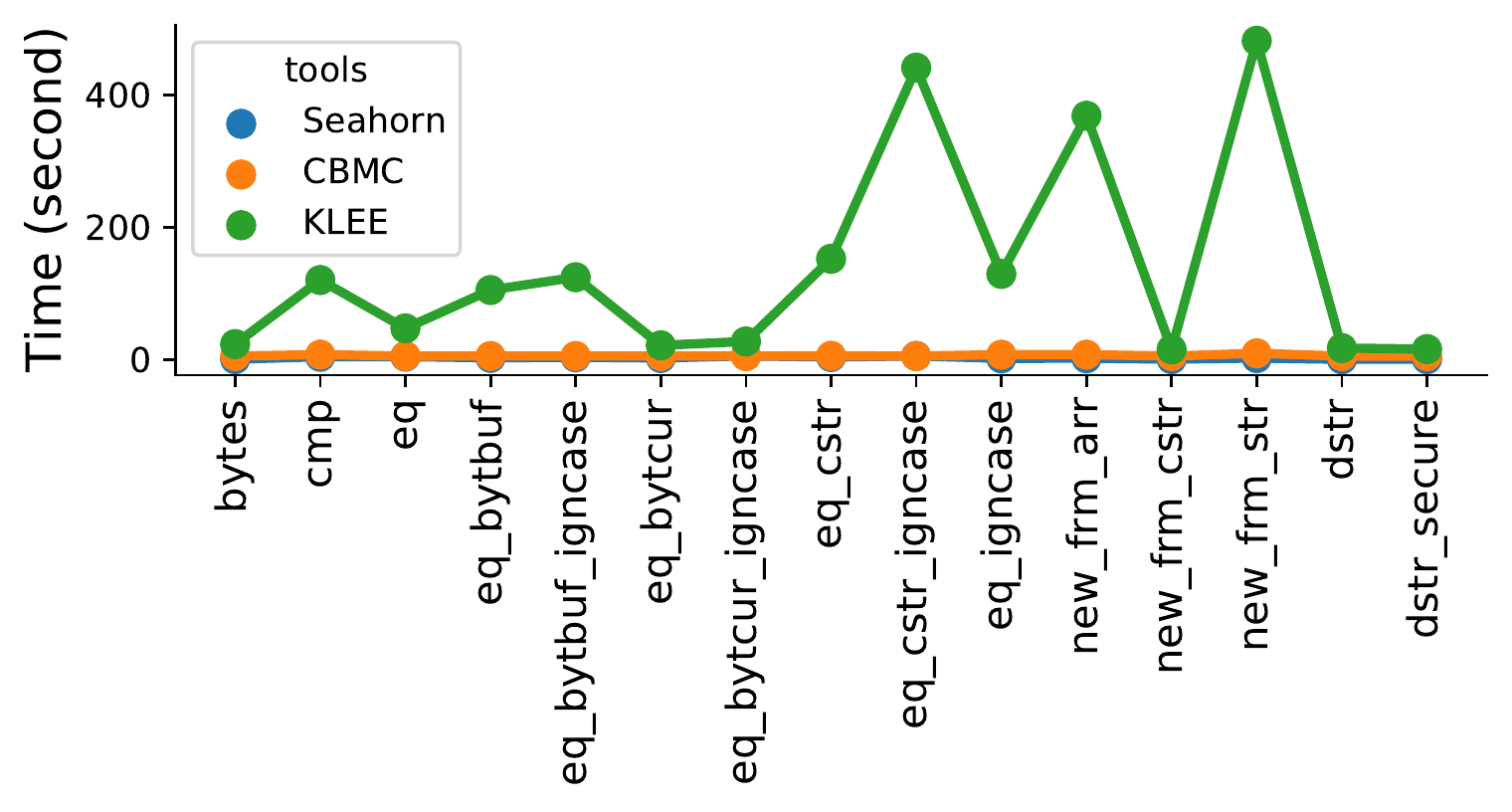}
  \caption{\code{aws_string}}
  \label{fig:comp_string}
\end{subfigure}\caption{Running times for \cbmc, \seahorn, and \klee on \awsccommon.}
\label{fig:comp_graphs}
\end{figure} \paragraph{Background.}  In the experiments, for \cbmc we use the original unit proof available at \awsccommon GitHub repository\footnote{https://github.com/awslabs/aws-c-common/tree/main/verification/cbmc/proofs.}. \seahorn and \klee use our adapted unit proofs and our implementation of verification library, including implementation of \code{nd_<type>} functions and initialization helpers. Obviously, \cbmc uses its internal compiler and pre-processing, while other tools rely on Clang compiler of LLVM. For \cbmc, we measure its total time. For \seahorn and \klee the compilation time is excluded because files are pre-compiled. The compilation time is insignificant. Finally, for \klee we had to severely restrict the sizes of allocated memory.  

For \libfuzzer, we do not show the running times because the most important result is coverage report. The details of workflow is shown in~\cref{fig:arch}. 

\vspace{0.1in}
\paragraph{Description.} The results are shown in ~\cref{fig:comp_graphs}. They are divided into categories. Each graph shows unit proofs on the $x$-axis and verification time on the $y$-axis. The category \emph{Hash} combines \code{hash_callback} and \code{hash_inter}, and  \emph{Others} combines \code{arithmetic} and \code{array}. 

Note that our performance comparison must be taken with a grain of salt. We have focused on developing a methodology to adapt the unit proofs to all the different tools. In some cases, this required significant change to what is being verified, both in terms of sizes of data structures considered, and in the properties checked (especially for built-in checks, such as memory safety). In our experience, all tools produced useful results. In particular, all the violations in the specifications have been found using \seahorn. However, these differences might have a significant effect on the running time. We leave a more in-depth study of performance of different techniques on the \emph{same} verification problems for future work.

  \fi

\end{document}